\documentclass[runningheads]{llncs}
\usepackage{framed}
\usepackage{multirow}
\usepackage{booktabs}
\usepackage{ifthen}
\usepackage{color}
\usepackage{url}
\usepackage{amsmath}
\usepackage{xspace}
\usepackage[T1]{fontenc}
\usepackage{enumerate}
\usepackage{array,multirow,graphicx}
\usepackage{float}
\usepackage{balance}
\usepackage{tikz}
\usepackage{calc}
\usepackage{subfigure}
\usepackage{listings,amsfonts}
\usepackage{amsmath,xcolor,pifont}
\usepackage{url}
\usepackage{xspace}
\usepackage{hyperref,endnotes}
\usepackage{xcolor}
\usepackage{array}
\usepackage{multirow,makecell}
\usepackage[misc]{ifsym}
\usepackage{tcolorbox}
\usepackage{pifont}
\usepackage{caption}
\usepackage{enumitem}
\usepackage{algorithm}
\usepackage[noend]{algpseudocode}

\begin{document}

\author{Tianyang Chi\inst{1} \and Ningyu He\inst{2} \and Xiaohui Hu\inst{3} \and Haoyu Wang*\inst{3}}

\institute{
Beijing University of Posts and Telecommunications, Beijing, China 
\and
The Hong Kong Polytechnic University, Hong Kong, China 
\and
Huazhong University of Science and Technology, Hubei, China 
}

\title{Remeasuring the Arbitrage and Sandwich Attacks of Maximal Extractable Value in Ethereum}
\maketitle

\begin{abstract}
Maximal Extractable Value (MEV) drives the prosperity of the blockchain ecosystem. By strategically including, excluding, or reordering transactions within blocks, block producers can extract additional value, which in turn incentivizes them to keep the decentralization of the whole blockchain platform.
Before September 2022, around \$675M was extracted in terms of MEV in Ethereum.
Despite its importance, current work on identifying MEV activities suffers from two limitations.
On the one hand, current methods heavily rely on clumsy heuristic rule-based patterns, leading to numerous false negatives or positives. On the other hand, the observations and conclusions are drawn from the early stage of Ethereum, which cannot be used as effective guiding principles after The Merge.
To address these challenges, in this work, we innovatively proposed a profitability identification algorithm. Based on this, we designed two robust algorithms to identify MEV activities on our collected largest-ever dataset.
Based on the identified results, we have characterized the overall landscape of the Ethereum MEV ecosystem, the impact the private transaction architectures bring in, and the adoption of back-running mechanisms.
Our research sheds light on future MEV-related work.
\end{abstract}

\section{Introduction}
Decentralized Finance (DeFi) has thrived in Ethereum~\cite{buterin2013ethereum}. In May 2024, DeFi reached a peak of \$109 billion in terms of Total Value Locked (TVL)~\cite{defillama}.
With the rise of DeFi participants, considerable attention has been garnered from the community since traders have identified multiple profitable opportunities, including capitalizing on price discrepancies of the same token across different DeFi projects.
Decentralized Exchanges (DEXes), as an important part of DeFi, are primarily responsible for the exchange of currency assets, which has created numerous such profit-making opportunities.
Typically, these opportunities depend on the transaction order on Ethereum, as they usually arise from other transactions interacting with these DEXes~\cite{daian2020flash}. Intuitively, traders must adjust the positions of their transactions within a block to seize profitable opportunities created by other transactions. This practice of profiting through reordering transactions is known as \textit{Maximal Extractable Value} (MEV)~\cite{daian2020flash,torres2021frontrunner,qin2022quantifying,wang2022cyclic,zhou2021high,qin2021empirical}.

Various transaction strategies and submission ways are currently adopted to conduct MEV. Attracted by the profitability of MEV, numerous accounts, widely known as \textit{MEV searchers}, are motivated to participate in MEV extraction.
Typically, there are two mainstream ways to perform MEV, \textit{i.e.,} \textit{arbitrage}~\cite{mclaughlin2023large,zhou2021just} and \textit{sandwich attacks}~\cite{zhou2021high,torres2021frontrunner}.
According to previous statistics~\cite{weintraub2022flash}, these two strategies have dominated the whole MEV ecosystem, accounting for more than 99\% of all MEV activities from May 4th, 2020 to March 23rd, 2022.

Current research on characterizing the MEV ecosystem is scant and limited.
On the one hand, \textit{their adopted methods are ineffective and inflexible.} Existing work widely adopts heuristic rule-based methods, not only leading to ineffective in identifying emerging advanced types of MEV activities, \textit{e.g.,} conjoined sandwich attacks, but also even ignoring whether the MEV initiators actually make a profit. 
On the other hand, \textit{their observations and conclusions are drawn from the early stage of Ethereum}. One of the milestones in Ethereum is \textit{The Merge}~\cite{ethereum_merge}, where the consensus algorithm is transitioned from \textit{Proof of Work} (PoW)~\cite{nakamoto2008bitcoin} to \textit{Proof of Stake} (PoS)~\cite{king2012ppcoin}. Moreover, \textit{Proposer-Builder Separation} (PBS) is subsequently introduced~\cite{pbs}, which distributes the block construction and verification responsibilities to different nodes. Only basic measurement work is conducted in this period~\cite{heimbach2023ethereum}, where more insightful conclusions are required.

To address these issues, this paper aims to conduct a comprehensive measurement study against the MEV ecosystem till the new era.
First, to the best of our knowledge, we have compiled the most extensive dataset to date, encompassing all transactions and blocks up to August 2023, both before and after The Merge and the introduction of PBS.
Then, we propose a profitability identification algorithm, which innovatively imports the concept of \textit{token exchange rate}. By calculating the exchange rate between any two tokens, the algorithm avoids the inaccuracies introduced by directly comparing token amounts.
Based on it, we further design two new algorithms to identify arbitrages and sandwich attacks, including both naive and advanced forms.
Finally, we conduct a comprehensive measurement to depict the whole MEV ecosystem in Ethereum, driven by the following three research questions:

\textit{RQ1: How is the effectiveness of our identification methodologies? What about the picture of MEV activities in Ethereum?}
To address this question, we apply our arbitrage and sandwich attack identification algorithms to the collected dataset, followed by a comparison with current state-of-the-art methods to quantitatively validate the effectiveness of our algorithms. Furthermore, we assess the involvement of DeFi smart contracts in MEV activities. (refer to \S\ref{sec:RQ1})

\textit{RQ2: What are the pros and cons of bringing private transaction architecture into the MEV ecosystem?}
The architecture for private transactions, designed to prevent MEV strategy exposure through transaction broadcasting, has experienced two significant restructurings within Ethereum. We categorize the timeline into three phases and analyze key metrics in each, including volume, financial metrics, and MEV strategies involved in MEV activities. By addressing these aspects, we aim to quantitatively demonstrate the positive and negative impacts that this new architecture has had on the community. (refer to \S\ref{sec:RQ2})

\textit{RQ3: What are the characteristics of the emerging back-running arbitrages?}
As competition among MEV searchers intensifies, back-running within the same block to extract MEV opportunities is increasingly becoming mainstream~\cite{mclaughlin2023large}. Consequently, we begin by quantifying existing back-running MEV activities. We then concentrate on two representative applications that utilize back-running arbitrages, \textit{i.e.,} MEV-Share/MEV-Block and Builder MEV extraction, to examine such an emerging type of MEV activity from various perspectives. (refer to \S\ref{sec:RQ3})

In this paper, against the three RQs, we have the following findings:
\begin{itemize}[leftmargin=*]
    \item Compared to state-of-the-art methods, our method can detect 16.4\% and 1.2\% more arbitrages and sandwich attacks, respectively, with at most 2.4\% false positive/negative rates. 
    \item Out of our collected largest-ever dataset, we identify 9.4 million in-the-wild MEV activities, where more than 100K DEX addresses and 129K tokens are involved. Except for well-known DeFi projects, like Uniswap V2, we observed that emerging meme tokens are becoming popular targets for MEV activities due to their volatile prices, smaller volumes, and greater liquidity.
    \item We conclude that the emergence of private transaction architecture pushes MEV searchers to abandon the mempool for conducting MEV activities. This shift is partly due to the low success rate (<40\%) of MEV execution via the mempool. Moreover, private transaction architecture avoids fees from failed transactions, turning previously unprofitable opportunities viable.
    \item Back-running arbitrages are gradually becoming a type of MEV activity that cannot be overlooked in both terms of frequency and revenue. However, the statistics on the adoption of MEV-Share and MEV-Blocker indicate back-running is far from fully utilized yet, while 620K USD are quietly and maliciously obtained by exploiting back-running mechanisms in Builders, urging a need for better regulatory protocols.
\end{itemize}

Our dataset will be released to the research community for further study.

\section{Background}
\subsection{Ethereum}
\label{sec:background:eth}
As one of the most well-known blockchain platforms, the killer application of Ethereum is smart contracts. 
All interactions among accounts, including smart contracts, are encoded in transactions, which are then packed into blocks.
For example, if accounts intend to initiate token transfers or invoke contracts, they will broadcast the transaction to the \textit{mempool}. \textit{Block creators}, who collect transactions and build blocks, select transactions from their mempools. As transactions in mempool can be observed by all nodes in Ethereum, some accounts prefer \textit{private transactions}, which are directly sent to the designated \textit{private transaction pool} owned by block creators.

In September 2022, Ethereum hits the milestone, \textit{The Merge}, indicating Ethereum 2.0~\cite{ethereum_merge}. One of the significant updates is the \textit{proposer-builder separation} (PBS)~\cite{pbs}. Specifically, \textit{block builders} are responsible for collecting and assembling transactions into a block body and then sending it to the relay, while \textit{block proposers} propose the block after signing the block header, sending it to the relay and accepting the full block constructed by the relay~\cite{heimbach2023ethereum}. Such a separation of duty improves the decentralization of Ethereum.

\subsection{Decentralized Finance \& Tokens}
\label{sec:background:defi}
Ethereum allows smart contracts to interact with each other for a complicated functionality, where they are jointly named as \textit{decentralized applications} (DApps)~\cite{dapp}.
More specifically, if a DApp serves a financial purpose, we often refer it to a \textit{decentralized finance} (DeFi) DApp.
DeFi DApps allow users to perform financial operations based on decentralized platforms instead of through the real-world centralized and regulated ones.
Currently, the DeFi DApps cover a lot of areas, \textit{e.g.,} currency exchange~\cite{lehar2021decentralized}, fund borrowing and lending~\cite{aavev1,aavev2,aavev3}, liquidity management~\cite{yearnfinance}, and cross chain~\cite{synapsebridge}.
Among all blockchain platforms, Ethereum has the most prosperous DeFi ecosystem. According to the statistics~\cite{defillama}, its TVL peaked in November 2021, around \$108 billion, around 60\% of all TVL among all public blockchain platforms.
Out of various types of DeFi DApps, \textit{decentralized exchange} 
 (DEX) is the most prevalent one.
Swap is the basic behavior on DEXes, through which users can exchange one type of token for another. Due to inadequate liquidity, users may encounter \textit{price slippage}~\cite{uniswap_slippage}, which refers to the price gap between the actual trading price and the quoted price at the time of transaction.

In the world of DeFi and DApps on Ethereum, \textit{tokens} serve as both a medium of exchange and a representation of value. Beyond Ethereum's own native token, Ether, a prevalent method for token creation is the ERC-20 standard~\cite{eip20}. This standard is critical because it provides a uniform set of interfaces, including functions like \texttt{transfer} and \texttt{approve}, which are essential for interacting with tokens in a standardized way across various applications.

\subsection{Maximal Extractable Value Extraction}
\label{sec:background:mev}
\textit{Maximal extractable value} (MEV) refers to any value that can be extracted from block production beyond the standard block reward and gas fees. MEV can be extracted by selecting, adjusting, or reordering transactions during processing a block~\cite{daian2020flash}.
For those accounts that figure out such opportunities and initiate MEV transactions, we name them as \textit{MEV searchers}.

\subsubsection{Front-running \& Back-running.}
\label{subsub:background:mev:front-running}
Reordering transactions is a key tactic in MEV extraction and its two main strategies are \textit{front-running} and \textit{back-running}. 
These strategies differ fundamentally in the positioning of the MEV searcher's transactions relative to the target transaction. In front-running, the MEV transaction precedes the target transaction, positioning it to directly influence the outcome or context of the target transaction by anticipating its effects. Conversely, in back-running, the MEV transaction is placed behind the target transaction, capitalizing on the market changes triggered by the target's execution.

\subsubsection{Common Types of MEV}
\label{subsub::three MEV types}
It is widely recognized that there are two mainstream types of MEV activities, \textit{i.e.,} \textit{arbitrage} and \textit{sandwich attack} ~\cite{qin2022quantifying,weintraub2022flash,wang2022cyclic,qin2021empirical,zhou2021high}.
Specifically, arbitrage refers to exploiting price differences between DEXes to achieve profits. For example, an MEV searcher swaps token $A$ for $B$ on $DEX_1$ and then swaps $B$ back for $A$ on $DEX_2$. Due to their price difference, the trader may get profits from such behaviors.
As for the sandwich attack, it refers to exploiting both price slippage (see \S\ref{sec:background:defi}) and transaction order to impact by back-running and front-running against one or several victim transactions. For example, on a specific DEX, a victim transaction intends to swap token $A$ for $B$. Before it is processed, an attacker front-runs the transaction by swapping $A$ for $B$, altering the exchange rate of the $A-B$ pair on the DEX. As a result, the victim receives fewer token $B$ than expected. Afterward, the attacker back-runs the transaction by swapping $B$ back for $A$ and profits from this price manipulation.

\begin{figure}[t]
    \centering
    \begin{minipage}{0.46\textwidth}
        \centering
        \includegraphics[width=\textwidth]{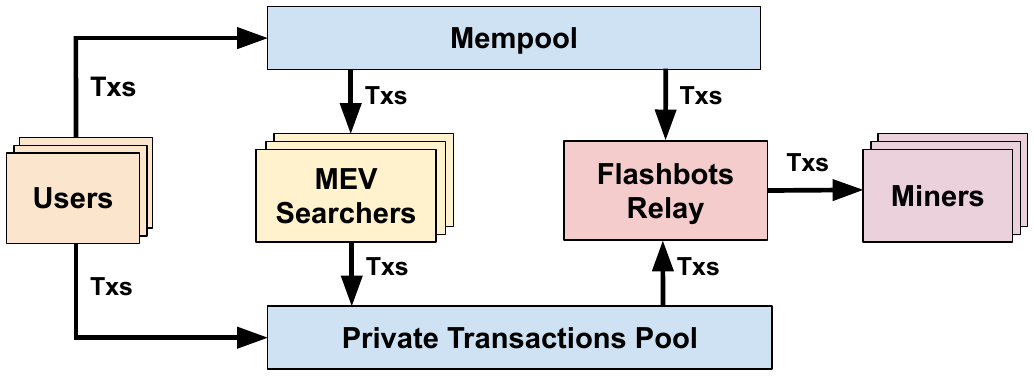}
        \vspace{-0.2in}
        \caption*{(a) Stage II}
        \label{fig:flashbots_a}
    \end{minipage}\hfill
    \begin{minipage}{0.51\textwidth}
        \centering
        \includegraphics[width=\textwidth]{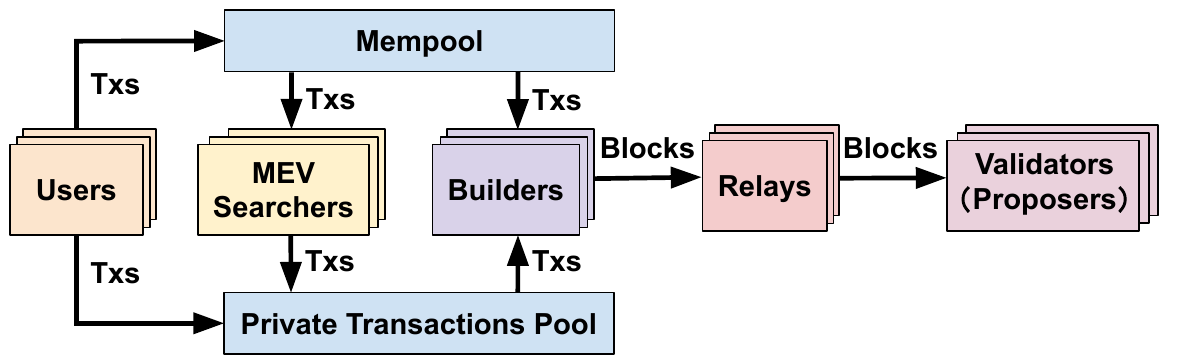}
        \vspace{-0.2in}
        \caption*{(b) Stage III}
        \label{fig:flashbots_b}
    \end{minipage}
    \vspace{-0.1in}
    \caption{Typical MEV architectures.}
    \vspace{-0.1in}
    \label{fig:flashbots}
\end{figure}

\subsubsection{MEV Forms}
\label{sec:background:mev:forms}
According to the involved infrastructures in MEV, we roughly divide the form of MEV into three stages:
\begin{itemize}[leftmargin=*]
\item \textbf{Stage I: Early-stage MEV}. When DeFi emerges, some MEV searchers figure out the opportunities of conducting arbitrages or sandwich attacks, thus they directly broadcast the MEV transaction into mempool. However, such transactions may be heard by other with conflicts with interest, who may imitate the heard transaction and initiate another one with a higher gas fee for a faster confirmation. This phenomenon is referred to as Priority Gas Auctions (PGAs)~\cite{daian2020flash}.
\item \textbf{Stage II: Flashbots-stage MEV}. The Stage I MEV may raise concerns about privacy and taking up space for normal transactions due to gas fee competition. Thus, Flashbots emerges to democratize MEV activities with private transaction pool. As we can see in Fig.~\ref{fig:flashbots}(a), Flashbots has a centralized relay, which accepts private transactions and MEV transactions directly from MEV searchers. Flashbots sorts transactions based on the profit searchers can get per unit of gas and packs them to miners, who are responsible for constructing blocks.
\item \textbf{Stage III: PBS-stage MEV}. Miners disappear when PBS emerges with Ethereum 2.0 (see \S\ref{sec:background:eth}). In addition, the centralized structure of Flashbots in Stage II leads to the concern about fairness. As shown in Fig.~\ref{fig:flashbots}(b), lots of \textit{relays} are responsible for relaying blocks constructed by builders to the final proposers. To this end, each relay adopts different strategies to choose blocks they received as proposers will choose the one that is most beneficial to them.  
\end{itemize}

\section{Data Overview}
\label{sec:data_overview}
To gain a comprehensive investigation of MEV activities, we need various types of data.
Therefore, we select different data sources to collect them, as shown in Table~\ref{tab:Dataset_Information}.
Firstly, we use the most well-known Ethereum client node, Geth, to sync transactions, emitted events, traces, and blocks.
In total, from block \#0 to \#18,000,000, we parsed over 2 billion Ethereum transactions, 3 billion emitted events and 5.5 billion traces.
Then, to filter out private transactions, we take adventage of \textit{Blocknative}~\cite{blocknative}, offering the most exhaustive historical archive of Ethereum's mempool transaction events. We take the ones that are included in Blocknative as private transactions, which strategy has also been adopted by other works~\cite{weintraub2022flash,qin2022quantifying}.
Moreover, for a more complete dataset of private transactions, we also query the \textit{Flashbots API}~\cite{flashbotsapi}, a public data source that records private transaction bundles passed through the Flashbots relays (see Fig.~\ref{fig:flashbots}(a)).
Consequently, 36 million private transactions are acquired.
Last, to collect all PBS blocks\footnote{PBS blocks cannot be distinguished according to block constructed time as there are still normal blocks after The Merge.}, we utilize the \textit{Relay API}~\cite{mevrelaylist}, a data source provided by each relay (see Fig.~\ref{fig:flashbots}(b)) consisting of blocks they have proposed.
Finally, we have collected more than 2 million PBS blocks.

\begin{table}[t]
  \centering
  \caption{Overview of our datasets.}
    \begin{tabular}{ccccc}
      \toprule
      \textbf{Source} & \textbf{Data Type} & \textbf{Covered Stages}* & \textbf{Time Span} & \textbf{Number} \\
      \midrule
      \multirow{4}{*}{\centering Geth} & Transaction & \multirow{4}{*}{Stage I/II/III} & \multirow{4}{*}{2015.07-2023.08} & 2,074,998,853 \\
      & Event & & & 3,108,423,890 \\
      & Trace & & & 5,582,166,611 \\
      & Block & & & 18,000,000 \\
      \hline
      Blocknative API & \multirow{2}{*}{Private Transaction} & \multirow{2}{*}{Stage II/III} & 2021.01-2023.08 & \multirow{2}{*}{36,180,823} \\
      Flahbots API & & & 2021.02-2022.09 & \\
      \hline
      Relay API & PBS block & Stage III & 2022.09-2023.08 & 2,087,069 \\
      \bottomrule
      \multicolumn{5}{l}{* Please refer to \S\ref{sec:background:mev:forms}}
    \end{tabular}
  \vspace{-0.15in}
  \label{tab:Dataset_Information}
\end{table}

\vspace{0.1in}
\noindent
\textbf{Ethical Consideration.}
This work collects several types of data, including Ethereum transactions from Etherscan, private transaction data from Blocknative API and Flashbots API, and PBS block data from Relay API.
Specifically, Ethereum data is accessible to everyone.
The data from the last three APIs are voluntarily provided by their respective organizations for research purposes.
We underline that all data collected in this work is publicly available and no ethical issues should be raised.

\section{Methodologies}
\label{sec:detection_algorithm}
Our study focuses on two types of MEV activities, \textit{i.e.,} arbitrage and sandwich attack. Previous works~\cite{wang2022cyclic,torres2021frontrunner,qin2022quantifying} widely adopt heuristic methods, which can lead to both false positives and false negatives according to our investigation.
We first demonstrate the limitations of previous methods with case studies \S\ref{sec:detection_challenge}. Then we propose our robust methods in \S\ref{sec:profitable_Identification} and \S\ref{sec:MEV_Identification}.

\subsection{Limitations of Existing Methods \& Case Studies}
\label{sec:detection_challenge}
\label{sec:case-Limitations}

According to our investigation on existing heuristic methods~\cite{wang2022cyclic,torres2021frontrunner,qin2022quantifying}, two main limitations on detecting arbitrages and sandwich attacks are discovered.

\noindent
\textbf{L1: Ineffectiveness in identifying profitable transactions.} 
Arbitrages and sandwich attacks typically require multiple swaps in DEXes, where previous methods directly compare the amount of tokens of the input and output of such swaps, leading to false positives in some cases.
For example, in the case of arbitrages, one heuristic rule is that the input amount of the first swap needs to be greater than the output of the final swap, ignoring that the token balance for the initiator may decrease. Thus, current heuristic method is ineffective to confirm whether the trader actually realizes a real-world profit.

\noindent
\textbf{L2: Insufficient flexibility of adopted rules.} 
Traditional identification rules lack flexibility and fail to adapt to the evolvement of the landscape of arbitrage and sandwich attacks. For example, in an arbitrage, each swap within the transaction may not necessarily be executed in chronological order, leading to false negatives in previous detection methods. Moreover, sandwich attacks may consist of more than two attack transactions. Previous methods, which limited the number of attack transactions to two, are unable to detect such attacks.

\label{sec:case-study}

\begin{figure}[t]
    \centering
    \begin{minipage}{0.49\textwidth}
        \centering
        \includegraphics[width=\textwidth]{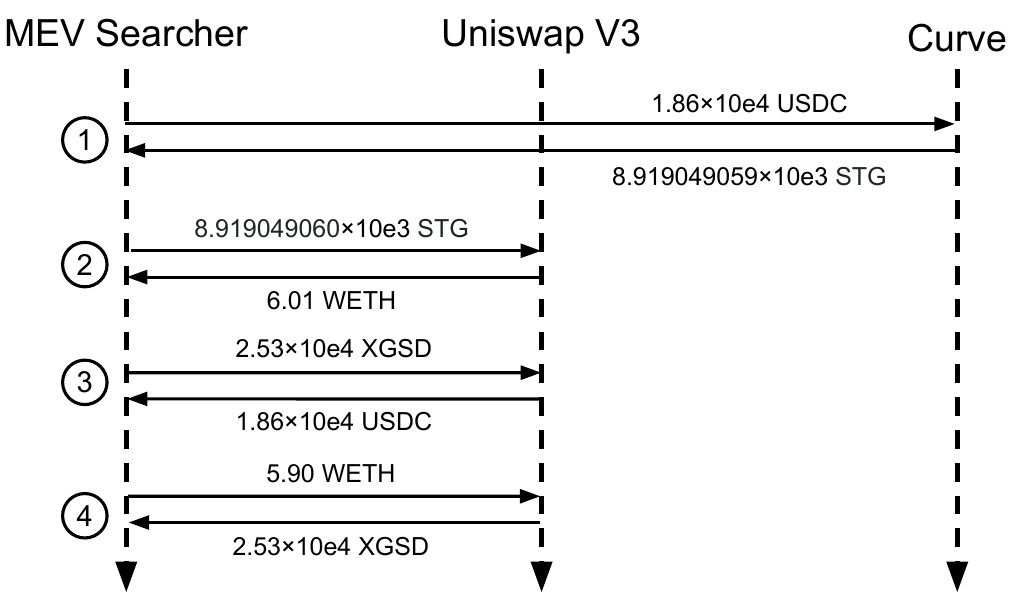}
        \vspace{-0.25in}
        \caption*{(a) Arbitrage}
    \end{minipage}\hfill
    \begin{minipage}{0.49\textwidth}
        \centering
        \includegraphics[width=\textwidth]{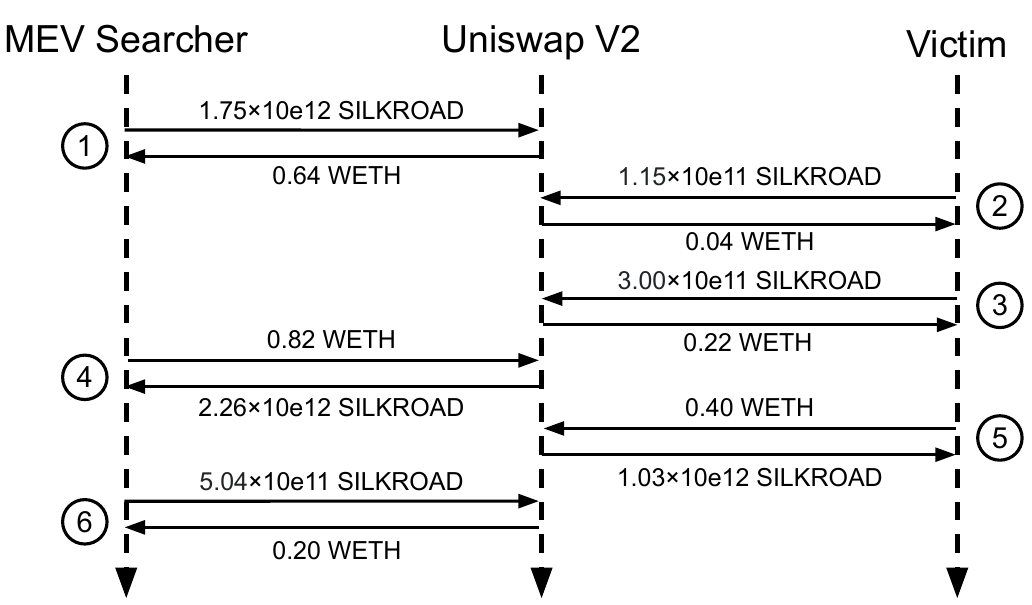}
        \vspace{-0.25in}
        \caption*{(b) Sandwich attack}
    \end{minipage}
    \vspace{-0.15in}
    \caption{Real-world MEV activities that cannot be identified by existing methods.}
    \vspace{-0.1in}
    \label{fig:arbitrage_sandwich_violation}
\end{figure}

We conduct two case studies to explain the limitations of existing methods.

\noindent
\textbf{Case 1: Arbitrage.}
Arbitrages are realized through multiple swaps. Current heuristic methods of identifying arbitrages are based on two rules: (1) the swaps should be conducted in sequential order, and (2) the output token quantity of the preceding swap should be at least equal to the input token quantity of the subsequent swap. However, these two intuitive rules can be violated, as illustrated in Fig.~\ref{fig:arbitrage_sandwich_violation}(a)\footnote{Transaction hash: 0x973cd5908d1ac430771688370f4c70405a887500abfe7d525ca215cd091d81d8}.
Firstly, due to the flashloan service, these four swaps do not form a circle in chronological order (the order of \ding{194} and \ding{195} is reversed).
Secondly, we can observe the input of \ding{193} is slightly greater than the output of \ding{192}, though this extra cost can be ignored compared to the final profit.
This example violates both heuristics but is still a successful arbitrage that will be taken as a false negative by previous methods.

\noindent
\textbf{Case 2: Sandwich Attack.}
In a typical sandwich attack, a victim transaction is flanked by two attack transactions, all involving token swap behaviors. Specifically, the first attack transaction and the victim transaction both swap token A for B, while the second attack transaction swaps token B back for A. The quantity of token B used in the second attack transaction should match the quantity obtained from the first attack transaction.
However, we have discovered a more complex form of sandwich attack, consisting of more than two attack transactions, as illustrated in Fig.~\ref{fig:arbitrage_sandwich_violation}(b).
As we can see, there are three attack transactions (\ding{192}, \ding{195}, and \ding{197}) and three victim transactions (\ding{193}, \ding{194}, and \ding{196}).
Moreover, we can observe that the amount of WETH between \ding{192} and \ding{195} and the amount of SILKROAD between \ding{195} and \ding{197} are different.
This example defies conventional expectations and remains undetectable by current methods, yet it is a successful sandwich attack since the attacker ultimately obtains additional profits in SILKROAD and WETH tokens.

\subsection{Profitability Identification}
\label{sec:profitable_Identification}

\begin{algorithm}[t]
    \caption{Profitability Identification Algorithm}
    \label{alg:check_profitable}
    \begin{algorithmic}[1]
        \Require \textit{tx} - a transaction or transactions; $g$ - a directed graph where nodes represent tokens and edges represent swaps.
        \Ensure A boolean value indicating if \textit{tx} is profitable for the trader.
        \State $\textit{addrBalChange} \leftarrow \text{getBalChange}(\textit{tx})$
        \State $\textit{addrBalChange} \leftarrow \text{rmvIrrAddr}(\textit{addrBalChange})$
        \State $\textit{tknChange} \leftarrow \text{aggrTknChange}(\textit{addrBalChange})$
        \For{$(\textit{tkn}_i, \textit{tkn}_k) $ \textbf{in} $\{(\textit{tkn}_i, \textit{tkn}_k) \mid \textit{tkn}_i,\textit{tkn}_k \in \textit{tknChange}$\ 
        $\land$ $tknChange.\textit{tkn}_i < 0$ $\land$  $\textit{tknChange}.\textit{tkn}_k > 0$ \}}
                \State $ratio \leftarrow (\textit{tkn}_i.out - \textit{tkn}_i.in) / \textit{tkn}_i.in$
                \If{$ratio < \epsilon$}
                \State  $exInput \leftarrow  -\textit{tknChange}.\textit{tkn}_i $
                \State  $exAmt \leftarrow \text{exchangeToken}(\textit{tkn}_i, \textit{tkn}_k, g, exInput)$
                    \If{$ \textit{tknChange}.\textit{tkn}_k - \textit{exAmt} > 0  $}
                        \State  $\textit{tknChange}.\textit{tkn}_k   \leftarrow   \textit{tknChange}.\textit{tkn}_k -\textit{exAmt} $ 
                        \State  $\textit{tknChange}.\textit{tkn}_i   \leftarrow 0$
                        
                    \EndIf
            \EndIf
        \EndFor
        
        \If{$\forall \text{tkn}_i \in \text{tknChange} (\text{tkn}_i \geq 0)$}
            \State \Return{\textbf{true}}
        \Else
            \State \Return{\textbf{false}}
        \EndIf
    \end{algorithmic}
\end{algorithm}

\begin{algorithm}[t]
    \caption{Exchange Token}
    \label{alg:exchange_token}
    \begin{algorithmic}[1]
        \Require \textit{inToken} - exchanged token; \textit{outToken} - output token; $g$ - the directed graph in Algorithm~\ref{alg:check_profitable}; \textit{exInput} - exchange amount.
        \Ensure \textit{exOutput} - output amount.
        \State $routeList \leftarrow \text{DFS}(\textit{inToken}, \textit{outToken}, g)$
        \State $\textit{exOutList} \leftarrow []$
        \For{$route_i $ \textbf{in} $\textit{routeList}$}
            \State $ exOutput \leftarrow \textit{exInput}$
            \For{$\textit{swap}_i $ \textbf{in} $\textit{route}_i $}
                \State $exchangeRate \leftarrow \textit{swap}_i.\textit{outputAmt} / \textit{swap}_i.\textit{inputAmt}$
                \State $\textit{exOutput} \leftarrow \textit{exOutput} * exchangeRate$
            \EndFor
            \State $\textit{exOutList}.\text{append}(\textit{exOutput})$
        \EndFor
        \State \Return{\Call{max}{\textit{exOutList}}}
    \end{algorithmic}
\end{algorithm}

Profitability is the gold standard for a successful MEV. However, previous methods are clumsy in determining whether a transaction is profitable as shown in \S\ref{sec:detection_challenge}.
To this end, we propose an algorithm, as shown in Algorithm~\ref{alg:check_profitable}.

Generally speaking, Algorithm~\ref{alg:check_profitable} examines if the given transaction is profitable to the initiator by assessing if the losses for a type of token can be covered by the profits generated by another type of token.
At L1 in Algorithm~\ref{alg:check_profitable}, \texttt{getBalChange} gets the balance changes for involved addresses by parsing ERC-20 transfer events and transaction traces, and stores them in \texttt{addrBalChange}, a two-level dictionary.
Then, at L2, \texttt{rmvIrrAddr} excludes irrelevant addresses, \textit{i.e.,} token contracts, DEX addresses, and the black hole address. 
After this, L3 aggregates token changes for remaining addresses, \textit{i.e.,} attackers, and keeps them in \texttt{tknChange}, a dictionary, where a positive amount signifies profit gained by the trader and vice versa.
The main body of the algorithm, \textit{i.e.,} loop at L4, aims to assess if the trader can profit from the transaction by evaluating whether the value of the tokens gained outweighs the value of the tokens lost. 
Specifically, this loop iterates \texttt{tknChange} to identify a tuple where $tkn_i < 0$ and $tkn_k > 0$, \textit{i.e.,} the trader loses a certain amount of $tkn_i$. At L5, we calculate a $ratio$ to quantify if the loss is small enough, \textit{i.e.,} less than a threshold $\epsilon$, which can be covered by the profits gained through $tkn_k$. If it does, $tkn_i$ is converted into an equivalent value of $tkn_k$, with corresponding updates to their balance changes (L9 -- L11).
The \texttt{exchangeToken} function that converts the number of lost $tkn_i$ into an equivalent value in $tkn_k$ at L8 is shown in Algorithm~\ref{alg:exchange_token}.
After several rounds of iteration, if all tokens in \texttt{tknChange} are non-negative, \textit{i.e.,} all losses are covered by profits, this indicates that the transaction (or series of transactions) is profitable for the trader (L12 -- L15).

Algorithm~\ref{alg:exchange_token} is responsible for exchanging a specific amount of one token for an equivalent value of another one by enumerating all possible swap paths.
Specifically, at L1, we traverse the directed graph to identify all acyclic paths from \textit{inToken} to \textit{outToken} in a DFS way, where each path represents a sequence of swaps.
The nested loop from L3 to L8 processes these paths, converting \textit{inToken} based on the derived exchange rates.
For example, if $tkn_i$ can be finally swapped to $tkn_k$ via $tkn_A$ and $tkn_B$, respectively, the amounts of swapped $tkn_k$ of two paths will be appended to $exOutList$.
After collecting all converted balances, the algorithm returns the maximum value of $exOutList$. This is because the $exInput$ is actually $-tknChange.tkn_i$ (see L7 in Algorithm~\ref{alg:check_profitable}), if the maximum amount of $tkn_k$ can still lead to profit, no matter which path is chosen, it can be guaranteed that the loss of $tkn_i$ is covered by the profit of $tkn_k$.

\subsection{Identifying Arbitrage \& Sandwich Attack}
\label{sec:MEV_Identification}
\begin{algorithm}[t]
    \caption{Identifying Arbitrages}
    \label{alg:identify_arbitrage}
    \begin{algorithmic}[1]
        \Require \textit{tx} - candidate transaction; \textit{swapPatterns} - collected swap events patterns.
        \Ensure A boolean value to indicate if the current transaction is an arbitrage, and a flag indicating whether this is a front- or back-running arbitrage.
        \State $swapList \leftarrow []$
        \For{$\textit{event}_i$ \textbf{in} $\textit{tx.events}$}
            \If{$\textit{event}_i \in swapPatterns$}
                \State $swapList.\text{append}(\text{swapParse}(\textit{event}_i))$
            \EndIf
        \EndFor
        \State $g \leftarrow \text{initGraph}(swapList)$
        \If{not $\text{existsCycle}(g)$}
            \State \Return{\textbf{false}, \textbf{null}}
        \EndIf
        \If{\text{checkProfitable(tx, g)}}
            \If{\text{replayTx(tx)}}
                \State \Return{\textbf{true}, \textbf{Front-running arbitrage}}
            \Else
                \State \Return{\textbf{true}, \textbf{Back-running arbitrage}}
            \EndIf
        \EndIf
        \State \Return{\textbf{false}, \textbf{null}}
    \end{algorithmic}
\end{algorithm}

\begin{algorithm}[t]
    \caption{Identifying Sandwich Attacks}
    \label{alg:detection_sandwich}
    \begin{algorithmic}[1]
        \Require \textit{txs} - candidate transactions; \textit{swapPatterns} - collected swap event patterns.
        \Ensure \textit{profitableAttack} - a list of sandwich attacks.
        \State initSwapList($txs$)

        \For{$\textit{tx}_k$ \textbf{in} $\textit{txs}$}
            \For{$\textit{event}_i$ \textbf{in} $\textit{tx}_k.\textit{events}$}
                \If{$\textit{event}_i \in swapPatterns$}
                    \State $\textit{tx}_k.swapList.\text{append}(\text{swapParse}(\textit{event}_i))$
                \EndIf
            \EndFor
        \EndFor

        \State $attackList \leftarrow []$
        \For{$\textit{tx}_i, \textit{tx}_v, \textit{tx}_k$ \textbf{in} $\textit{txs}$}
            \If{\text{isChrono}($\textit{tx}_i, \textit{tx}_v, \textit{tx}_k$) $\land$ 
                $(\textit{tx}_i.\textit{from} = \textit{tx}_k.\textit{from} \vee \textit{tx}_i.\textit{to} = \textit{tx}_k.\textit{to})$ $\land$ \text{checkSwapDirection}$(\textit{tx}_i,\textit{tx}_v,\textit{tx}_k)$
                }
                \State $\textit{attackList}.\text{append}((\textit{tx}_i, \textit{tx}_k))$
            \EndIf
        \EndFor
        \State $\textit{possibleAttack} \leftarrow \text{combineAttacks}(\textit{attackList})$
        \State $\textit{profitableAttack} \leftarrow []$
        \For{$\textit{attack} $ \textbf{in} $\textit{possibleAttack}$}
        \State $swaps \leftarrow \text{concatenate}([tx.swapList \text{ for } tx \in attack])$  
        \State $g \leftarrow \text{initGraph}(swaps)$
            \If{$\text{checkProfitable}(\textit{attack},g)$}
                \State $\textit{profitableAttack}$.\text{append}$(\textit{attack})$
            \EndIf
        \EndFor
        \State \Return{$profitableAttack$}
    \end{algorithmic}
\end{algorithm}

\subsubsection{Identifying Arbitrages.}
To comprehensively identify arbitrages and overcome the second limitation discussed in \S\ref{sec:case-Limitations}, we propose a new algorithm as shown in Algorithm~\ref{alg:identify_arbitrage}. It enhances circle detection without requiring sequential orders of swaps, and adopts Algorithm~\ref{alg:check_profitable} to determine the profitability for involved addresses.
Specifically, given a transaction, $tx$, the algorithm first parses all emitted events. Note that, we have summarized a $swapPatterns$, consisting of typical swap events used by DEXes\footnote{We extend existing works~\cite{weintraub2022flash,qin2022quantifying} by covering 44 different types of swap patterns, increasing 6.3x and 5.3x, respectively}.
In the loop at L2, if an event conforms to one in $swapPatterns$, it extracts key information including the input and output token contract, as well as their corresponding amounts, as a tuple in the $swapList$.
Using collected swaps, \texttt{initGraph} at L5 constructs a directed graph with nodes as tokens and edges as swaps, each edge representing the flow from input to output token with embedded amounts. If no cycles are detected (L6 -- L7), the transaction is not considered as an arbitrage considering the essence of arbitrages.
The \texttt{checkProfitable} function at L8 assesses whether the transaction yields profits for traders, as outlined in Algorithm~\ref{alg:check_profitable}. Those profitable transactions are classified as arbitrages.
As described in \S\ref{subsub:background:mev:front-running}, two types of MEV exist. Following McLaughlin's heuristic~\cite{mclaughlin2023large}, a transaction is classified as front-running if it remains an arbitrage when replayed at the block's start; otherwise, it is back-running arbitrage (L9 -- L12).

\subsubsection{Identifying Sandwich Attacks.}
The typical assumption of sandwich attack is that only one victim transaction exists. However, we find advanced sandwich attacks where multiple victim transactions are sandwiched between two attack transactions. We refer to this pattern as \textit{multi-layered burger attack}~\cite{li2023demystifying}. Additionally, if a sandwich attack involves multiple attack transactions like the one shown in the second case in \S\ref{sec:case-study}, we term it a \textit{conjoined sandwich attack}.

Algorithm~\ref{alg:detection_sandwich} details the method for detecting sandwich attacks. It starts by parsing all events from each transaction within the same block to identify swaps (L1 -- L5).
From L7 to L9, the algorithm searches for potential pairs of attack transactions according to the following three predicates. (1) it requires chronological relationships among them, where the victim one is in the middle; (2) the two potential attack transactions must either originate or target the same address; and (3) the swap directions of the two potential attack transactions should be opposite, while the victim one shares the same swap direction as the first potential attack transaction (ensured by \texttt{checkSwapDirection}).
At L10, the algorithm will aggregate all feasible attack transaction pairs. For example, if a transaction is involved in two attacks, they will be combined and taken as a \textit{conjoined sandwich attacks}.
From L12 to L16, the algorithm iterates all combinations, identifying and storing potentially profitable ones in $profitableAttack$.

\section{Study Design}
\label{sec:study-design}

\paragraph{Research Questions.}
With our proposed methods, we raise the following research questions (RQs).

\begin{itemize}
	\item[RQ1] How is the effectiveness of our identification methodologies? What about the picture of MEV activities in Ethereum? 
	\item[RQ2] What are the pros and cons of bringing private transaction architecture into the MEV ecosystem? 
	\item[RQ3] What are the characteristics of the emerging back-running arbitrages?
\end{itemize}

\paragraph{Dataset Comparison.}
\label{sec:eval:comparison}
After a comprehensive literature review to the best of our efforts, we found 8 studies~\cite{li2023demystifying,mclaughlin2023large,park2024unraveling,piet2022extracting,qin2022quantifying,torres2021frontrunner,wang2022cyclic,wu2021defiranger} are related to detecting arbitrages and sandwich attacks.
As for arbitrage MEV activities, among these, only Weintraub~\cite{weintraub2022flash} has open-sourced the corresponding dataset. Although Z. Li \textit{et al.}~\cite{li2023demystifying} represent the state-of-the-art method according to their experimental results, we do not take their results as the baseline as the lack of a publicly available dataset.
As for MEV sandwich attacks, we found all previous studies~\cite{heimbach2023ethereum,heimbach2024non,park2024unraveling,yang2022sok} used the ZeroMEV dataset~\cite{ZEROMEV}.
Consequently, we take two available MEV datasets for comparison, \textit{i.e.,} Weintraub~\cite{weintraub2022flash} for arbitrages, and ZeroMEV~\cite{ZEROMEV} for sandwich attacks.

\paragraph{Experimental Setting.}
\label{sec:eval:setting}
While previous research has concentrated on only a few DEXes, we aim to identify MEV activities across a broad range of DEXes. To achieve this, we review all DEXes listed on DefiLlama~\cite{defillama}, analyzing the swap patterns they include and applying our identification methods. As a result, we cover 44 swap patterns in our analysis.
To answer RQ1, we manually investigate their implementations and extract their swap event patterns, which are required by our methods (see Algorithm~\ref{alg:identify_arbitrage} and Algorithm~\ref{alg:detection_sandwich}).
To answer RQ2 and RQ3, we need to set up a replay environment.
By forking the code of Ganache~\cite{ganache}, we can replay any transaction at any position.

\section{RQ1: Status of MEV Activities}
\label{sec:RQ1}
In this section, we apply our methodologies to identify MEV activities and compare their effectiveness with existing ones. We also quantify what roles are played by DEXes and tokens in the MEV ecosystem.

\subsection{Identified MEV Activities}

\begin{figure}[t]
\begin{minipage}{0.4\textwidth}
\captionof{table}{Overview of identified MEV activities.}
\label{tab:overview_MEV}
\resizebox{\textwidth}{!}{
\begin{tabular}{@{}lr@{}}
    \toprule
    \textbf{Type} & \textbf{Count} \\ 
    \midrule
    Arbitrages & 6,331,851\\
    Sandwich Attacks & 3,016,971 \\
    \quad Multi-layered Burger Attacks & 395,779 \\
    \quad Conjoined Sandwich Attacks & 31,878 \\
    \quad Normal Sandwich Attacks& 2,589,314 \\
    \midrule
    \textbf{Total} & 9,348,822 \\
    \bottomrule
    \end{tabular}
}
\end{minipage}
\hspace{3pt}
\begin{minipage}{0.48\textwidth}
    \centering
    \captionof{table}{Reasons of exclusively identified arbitrages in our method compared to Weintraub.}
\label{tab:arb_reasons}
\resizebox{\textwidth}{!}{
    \begin{tabular}{lrr}
    \toprule
    \textbf{Reasons} & \textbf{Count} & \textbf{\%} \\
    \midrule
        Limited swap patterns & 566,076 & 81.6\% \\ 
        Strict token amount comparison & 76,506 & 11.0\% \\
        Strict chronological examination & 26,539 & 3.8\% \\
        Other & 24,669 & 3.6\% \\ 
    \midrule
        \textbf{Total} & \textbf{693,790} & \textbf{100.0\%} \\ 
    \bottomrule
  \end{tabular}
}
\end{minipage}
\end{figure}

\begin{figure}[t]
    \centering
    \begin{minipage}{0.49\textwidth}
        \centering
        \includegraphics[width=\textwidth]{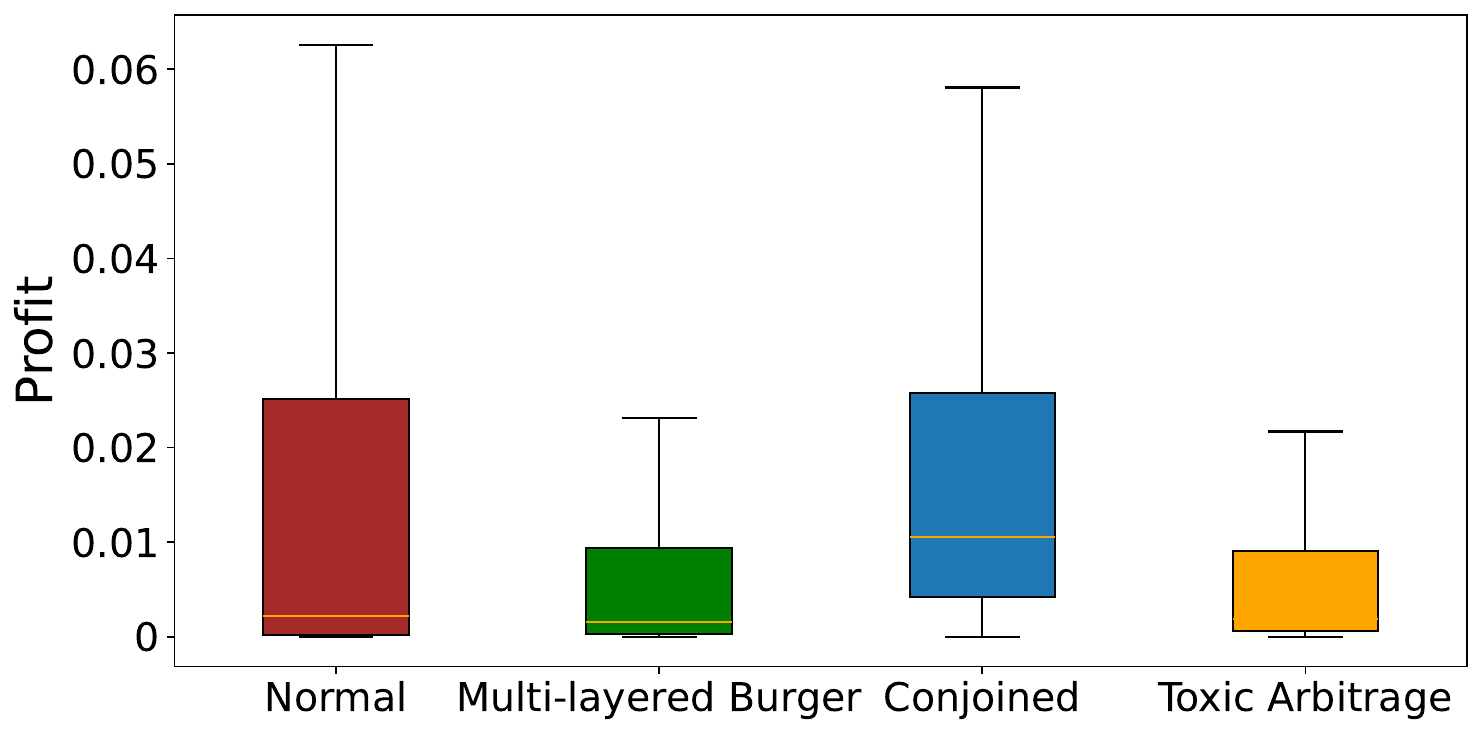}
        \caption*{(a) Profit}
    \end{minipage}\hfill
    \begin{minipage}{0.49\textwidth}
        \centering
        \includegraphics[width=\textwidth]{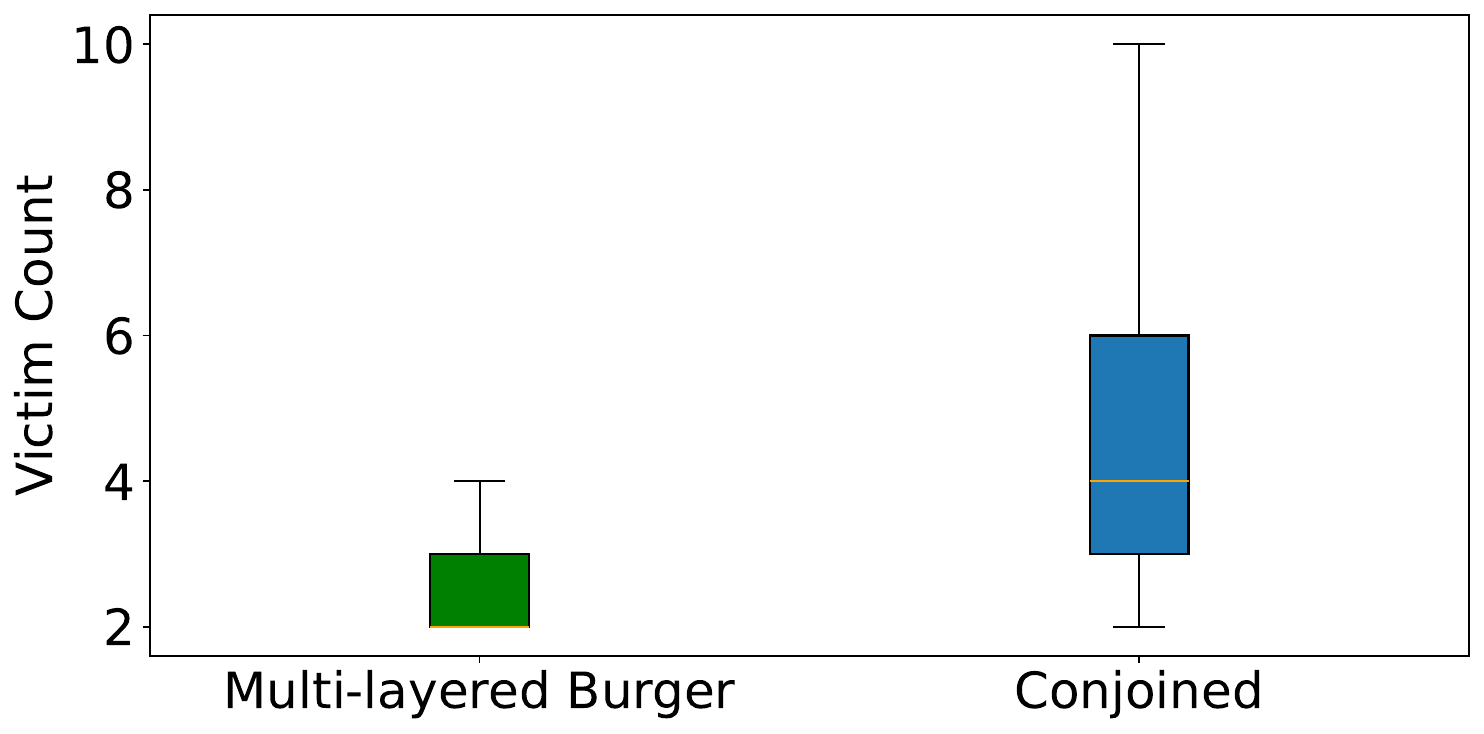}
        \caption*{(b) Transaction Count}
    \end{minipage}
    \vspace{-0.1in}
    \caption{Profit and transaction count for different types of sandwich attacks.}
    \vspace{-0.1in}
    \label{fig:sandwich_victim_and_profits}
\end{figure}

As detailed in Table~\ref{tab:overview_MEV}, we have successfully identified approximately 9.4 million MEV activities, comprising 6,331,851 arbitrages and 3,016,971 sandwich attacks. Notably, we discovered an overlap between these two categories, involving 148,133 transactions.
Upon analyzing the behavioral patterns of these overlapping cases, we categorized them as \textit{toxic arbitrages}~\cite{toxicarbitrage2022}, referring to a type of sandwich attack in which one or both attack transactions simultaneously qualify as profitable arbitrages.
In addition, as we stated in \S\ref{sec:MEV_Identification}, our method can identify some advanced sandwich attacks. In total, there are 395,779 (13.1\%) multi-layered burger attacks and 31,878 (1.1\%) conjoined sandwich attacks (see \S\ref{sec:MEV_Identification}).
To depict their characteristics, we first compare the profits among these sandwich attacks, as shown in Fig.~\ref{fig:sandwich_victim_and_profits}(a). We found that the median profits of multi-layered burger attacks, toxic arbitrages and normal sandwich attacks are quite similar, while the profits of conjoined sandwich attacks are almost 5 times than others.
We found that the reason is that the attack principles of sandwich attacks and multi-layered burger attacks are similar, but in conjoined sandwich attacks, attackers can use multiple transactions to cause a larger price deviation, thereby obtaining more profits from the victim transaction.
As both advanced sandwich attacks can hold more than one victim transaction, we then analyze the number of victim transactions, as shown in Fig.~\ref{fig:sandwich_victim_and_profits}(b).
We can observe that the median number is 2 and 4, respectively.
Involving more victim transactions could further increase the profits of conjoined sandwich attacks.

\textbf{\textit{Findings:}} Out of 2 billion transactions, we identified near 9.4 million MEV activities. Moreover, we can easily observe that complicated MEV activities are significantly fewer in number but can generate higher profits, raising challenges but also providing opportunities to MEV searchers.

\subsection{Comparison with Existing Methods}
\label{sec:method_comparision}
To evaluate the effectiveness of our proposed methods, in this part, we compare our identification results with two datasets as mentioned in \S\ref{sec:eval:comparison}\footnote{For a fair comparison, we only compare the results over the same block range.}.

\subsubsection{Arbitrages Comparison.}
Within the given range, our method identifies a total of 3,910,444 arbitrages, whereas Weintraub can identify 3,358,371 ones\footnote{Both of them exclude the toxic arbitrage.}. 
Specifically, 3,216,654 transactions are detected by both methods, while 141,717 and 693,790 transactions are exclusively for Weintraub and our method, respectively.
To figure out the root causes of such distinctions, we sample five hundred transactions from our exclusively detected 693K arbitrages. We summarized three root causes, where the breakdown is shown in Table~\ref{tab:arb_reasons}.
Specifically, \textit{limited swap patterns} (81.6\%) refers to the scope of swap event patterns of DEXes is limited.
\textit{Strict token amount comparison} (11.0\%) and \textit{strict chronological examination} (3.8\%) respectively refer to the two limitations mentioned in the first case study in \S\ref{sec:case-study}. Unfortunately, for the remaining 3.6\%, we cannot identify the reason.
Though the limited swap patterns nearly dominate, we still cannot underestimate the impact brought by the two seemingly intuitive rules, around 15\% of transactions are mislabeled as negatives by Weintraub.

As for the 141K transactions that are not identified by our method, following the same method, we summarize two root causes.
One is that our method excludes those transactions with final token losses, accounting for 38,729 transactions (27.3\%).
Another is that Weintraub's method indiscriminately labels transactions with swap loops within a single DEX as arbitrages, accounting for 11,008 transactions (7.8\%). 
For the remaining 91,980 transactions, the sampling results show that they do obtain profits, which, however, are transferred to other entities, where some of them are labeled as \textit{MEV Bot} by Etherscan.
From the perspective of blockchain anonymity, our approach conservatively does not consider the case of collusion.

As for the false positive rate (FPR) and false negative rate (FNR), since we cannot determine the reasons for the 24,669 newly identified ones in Table~\ref{tab:arb_reasons}, we consider all of them as FPs as an upper bound.
Additionally, because 91,980 transactions are not detected by our method as we do not take collusion into consideration, we take all of them as FNs as an upper bound. 
Consequently, the FPR and FNR in detecting arbitrages are 0.6\% and 2.4\%, respectively.

\subsubsection{Sandwich Attack Comparison.}
Within the given block range, our method detects a total of 1,265,929 sandwich attacks, while ZeroMEV labels 1,251,072 ones.
Among them, 84,944 and 70,088 sandwich attacks are exclusively marked by our method and the ZeroMEV's, respectively.
Similarly, we sample 100 out of 70,088 sandwich attacks exclusively marked by ZeroMEV.
We find that 57 are non-sandwich attacks, involving various EOA addresses interacting with public contracts, like Uniswap V2 Router, with no evidence indicating that they are sandwich attacks.
Moreover, we speculate that 19 of them are likely sandwich attacks, but they have interacted with different addresses. Again, conservatively, we cannot ensure if they are profitable sandwich attacks.
Last, the remaining 24 transactions all result in token losses for the original transaction initiators, which are excluded by our method as we still stand behind the goal of MEV activities is to earn profits.
As for the ones that are exclusive by our method, we sample 100 of them and manually check them. We find that all of them are indeed sandwich attacks.

We also estimate the FPR and FNR of the sandwich attack detection method.
As we sample 100 transactions from the 70,088 sandwich attacks exclusively marked by ZeroMEV, we estimate that 43\% of these transactions are sandwich attacks, serving as an upper bound. The FNR is calculated as \(\frac{70,088 \times 43\%}{1,265,929}=2.4\%\).
As for FPR, since we do not find any FPs in the sampling that are exclusive to our method, our method has a negligible FPR compared to ZeroMEV.

\textbf{\textit{Findings:}} Compared to existing methods, our method outperforms both of them in identifying arbitrage and sandwich attacks. Under the upper-bound estimation, the FPR and FNR for arbitrage is 0.6\% and 2.4\%, while these two numbers for sandwich attacks are $\sim$0\% and 2.4\%.

\subsection{DEX \& Token Composition}
\label{sec:dex_composition}
We further analyze the DEXes and tokens involved in MEV transactions to better understand the MEV ecosystem.
Specifically, as we stated in \S\ref{sec:eval:setting}, we have collected 44 patterns of swap events, belonging to different mainstream DEXes. We collected a total of 30,659,911 swap events originating from MEV transactions: 17,300,465 (56.4\%) from arbitrages, 7,875,793 (25.7\%) and 5,483,653 (17.9\%) respectively from the attack and victim transaction of sandwich attacks.
Based on them, 
we have parsed 100,709 DEX and 129,909 token addresses.  

\begin{figure}[t]
    \centering
    \begin{minipage}{0.39\textwidth}
        \centering
        \includegraphics[width=\textwidth]{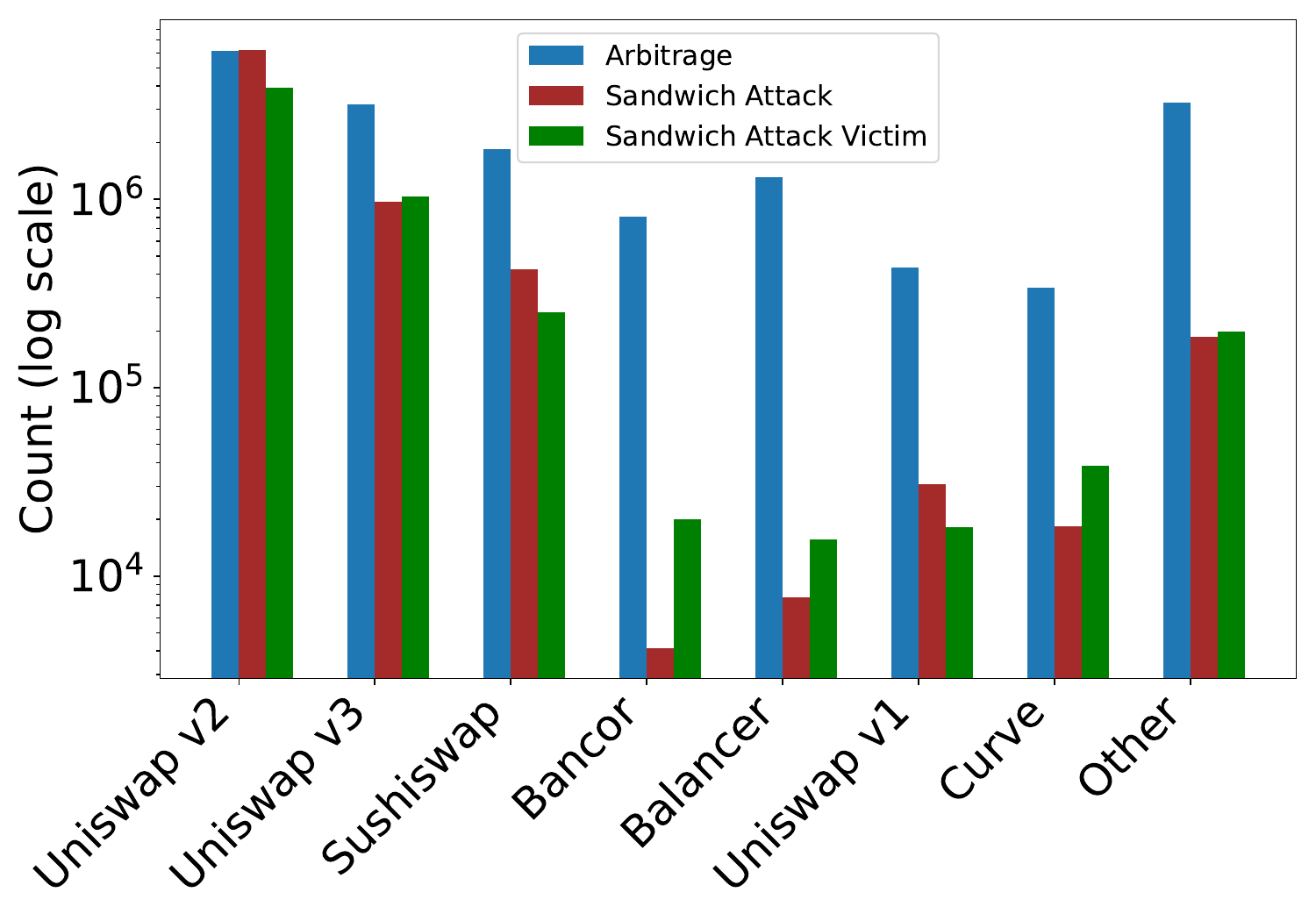}
        \vspace{-0.2in}
        \caption*{(a) Distribution of swap actions in different DEXes.}
    \end{minipage}\hfill
    \begin{minipage}{0.59\textwidth}
        \centering
        \includegraphics[width=\textwidth]{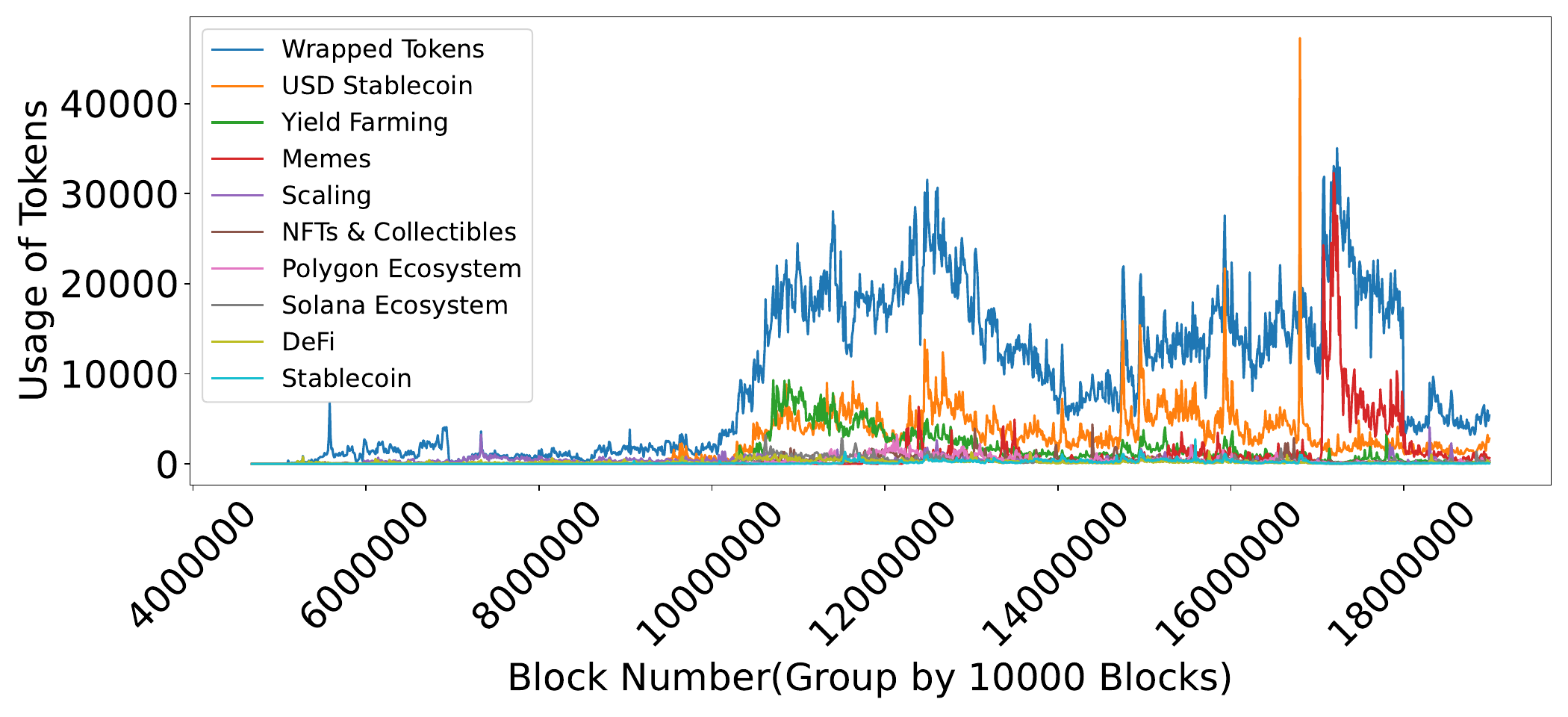}
        \vspace{-0.2in}
        \caption*{(b) Top 10 categories of tokens involved in MEV transactions.}
    \end{minipage}
    \vspace{-0.1in}
    \caption{Statistics of DEXes and token categories in MEV transactions.}
    \vspace{-0.1in}
    \label{fig:swap_count_and_token}
\end{figure}

Fig.~\ref{fig:swap_count_and_token}(a) shows which DEX is responsible for emitted swap events.
As we can see, Uniswap V2 dominates, accounting for 53.0\% of all emitted swap events.
These swap events are related to 81,818 DEX addresses, accounting for 81.0\% of all involved ones. In other words, more than 81K pools maintained by Uniswap V2 have been targets of MEV activities. 
Thus, traders on Uniswap V2 need to be vigilant and consider using private architectures, such as Flashbots, to cover their trade intents from MEV searchers.

For the 130K token addresses involved in MEV activities, we categorize them by the labeling data from CoinMarketCap~\cite{coinmarketcap}, a well-known cryptocurrency browser, to understand the degree of participation of different types of tokens in MEV transactions.
Fig.~\ref{fig:swap_count_and_token}(b) shows the participation of the top 10 categories in MEV transactions. As we can see, wrapped token~\cite{wraptoken}, like WETH and WBTC, is the most involved one, accounting for 97.0\% of MEV transactions. Following is USD stablecoins, such as USDC and USDT, participating in 22.9\% of MEV transactions. 
This observation is consistent with previous research~\cite{mclaughlin2023large,wang2022cyclic}, indicating that MEV transactions are closely related to such stablecoins.
However, we find that Meme tokens, like DOGE~\cite{DOGE}, have surpassed USD stablecoin to become the second-largest category involved in MEV transactions after April 16th, 2023. We believe there are two main reasons.
First, the trading hype of Meme tokens such as PEPE~\cite{pepecoin} has led to significant price fluctuations. Second, the relatively small liquidity pools of Meme tokens on DEXes lead to significant price slippage after trades. Both create MEV opportunities for MEV searchers.

\textbf{\textit{Findings:}} Due to the huge volume of well-known DEXes and tokens, they are popular targets for MEV activities. However, we have observed that emerging tokens, because of their frequent price fluctuations and possible price slippage, are also becoming targets of MEV activities. 

\begin{tcolorbox}[colback=white, colframe=black, sharp corners, boxrule=0.2mm, width=\linewidth]
\textbf{Answer to RQ1:}
Our identification methods against MEV activities outperform the state-of-the-art ones in terms of precision and recall. Currently, MEV activities that require meticulous construction and target emerging tokens are becoming the priority for MEV searchers, because they are likely to generate greater profits.
\end{tcolorbox}

\section{RQ2: Private MEV Activities vs. Mempool MEV Activities}
\label{sec:RQ2}
In this section, we aim to depict the distinction between MEV activities conducted in the mempool and private transaction architecture. Thus, we focus on the volume, financial metrics, and adopted strategies of MEV activities.

\subsection{Volume of MEV Activities}
\label{sec:rq2:tx-volume}

\begin{figure*}[t]
    \centering
    \includegraphics[scale=0.2]{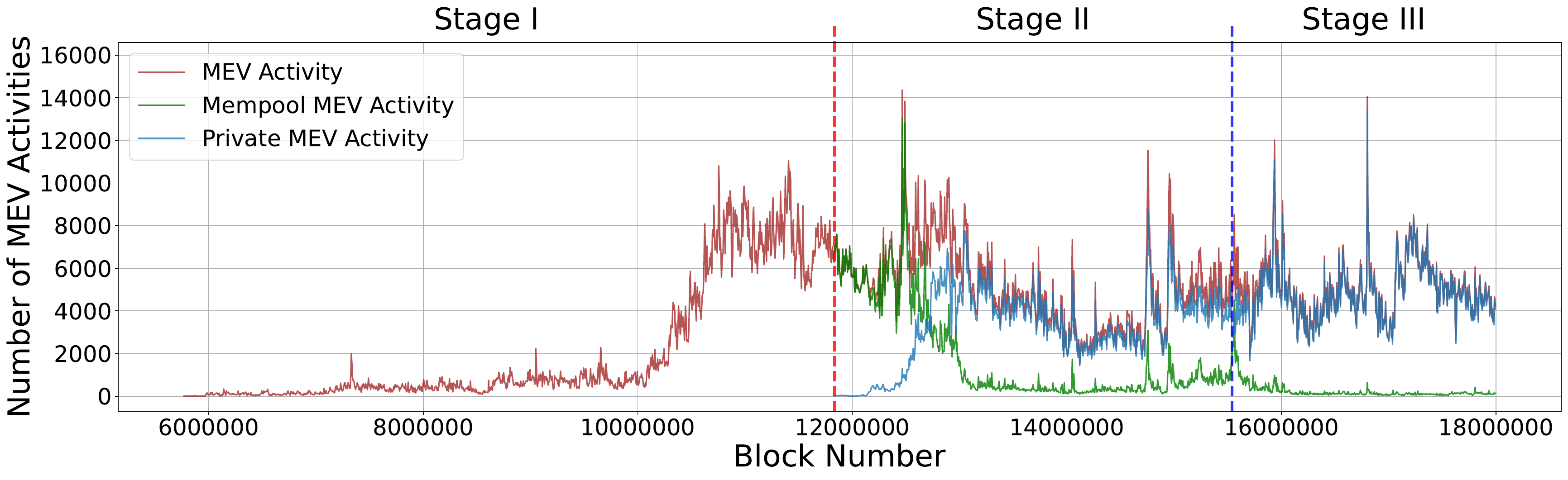}  
    \vspace{-0.1in}
    \caption{The number of MEV activities in different stages.}
    \vspace{-0.1in}
    \label{fig:Private_Pool}
\end{figure*}

We depict the number of MEV activities, as the red line shown in  Fig.~\ref{fig:Private_Pool}.
In Stage I, we can observe an obvious rapid growth, spanning from May 2020 to October 2020, which is attributed to the outburst of the Ethereum DeFi ecosystem. In total, in Stage I, there are 2,488,789 MEV activities.
Since Stage II, we can distinguish mempool and private MEV activities. We can observe a sharp decline in the number of mempool MEV activities, and it is consistently wiggling at a low level. Instead, the private transaction pool immediately dominated a significant portion of the MEV market.
We speculate the reasons for such a shift are twofold: (1) private MEV activities with higher transaction fees are prioritized and take MEV opportunities from mempool MEV activities (refer to \S\ref{sec:MEV profit}); and (2) mempool MEV activities have lower success rate, forcing MEV searchers to use private transactions (refer to \S\ref{sec:Success Rate}).
Although the number of mempool MEV activities slightly increased during the upgrade phase of the private transaction architecture, particularly around the boundary between Stage II and Stage III, the number of mempool MEV activities still rapidly decreased after the upgrade was completed.
Our statistics show that in Stage II, 37.1\% of MEV activities originated from the mempool, while the number went to only 4.7\% in Stage III.

\textit{\textbf{Findings:}} Once the private transaction architecture has emerged, including the centralized Flashbots Relay and the decentralized Relays, due to its processing priority over mempool MEV activities, MEV activities via private transaction pools started to dominate the whole ecosystem.

\subsection{Financial Metrics of MEV Activities}
\label{sec:MEV profit}
We do not consider MEV activities in Stage I as there lacks a widely-recognized source for private transactions in this stage. Thus, for Stage II and III, we quantitatively analyze the profit, profit margin, and revenue of MEV activities, providing a picture of how PBS impacts the profit distribution in the MEV ecosystem.

\begin{figure}[t]
    \centering
        \begin{minipage}{0.32\textwidth}
        \centering
        \includegraphics[width=\textwidth]{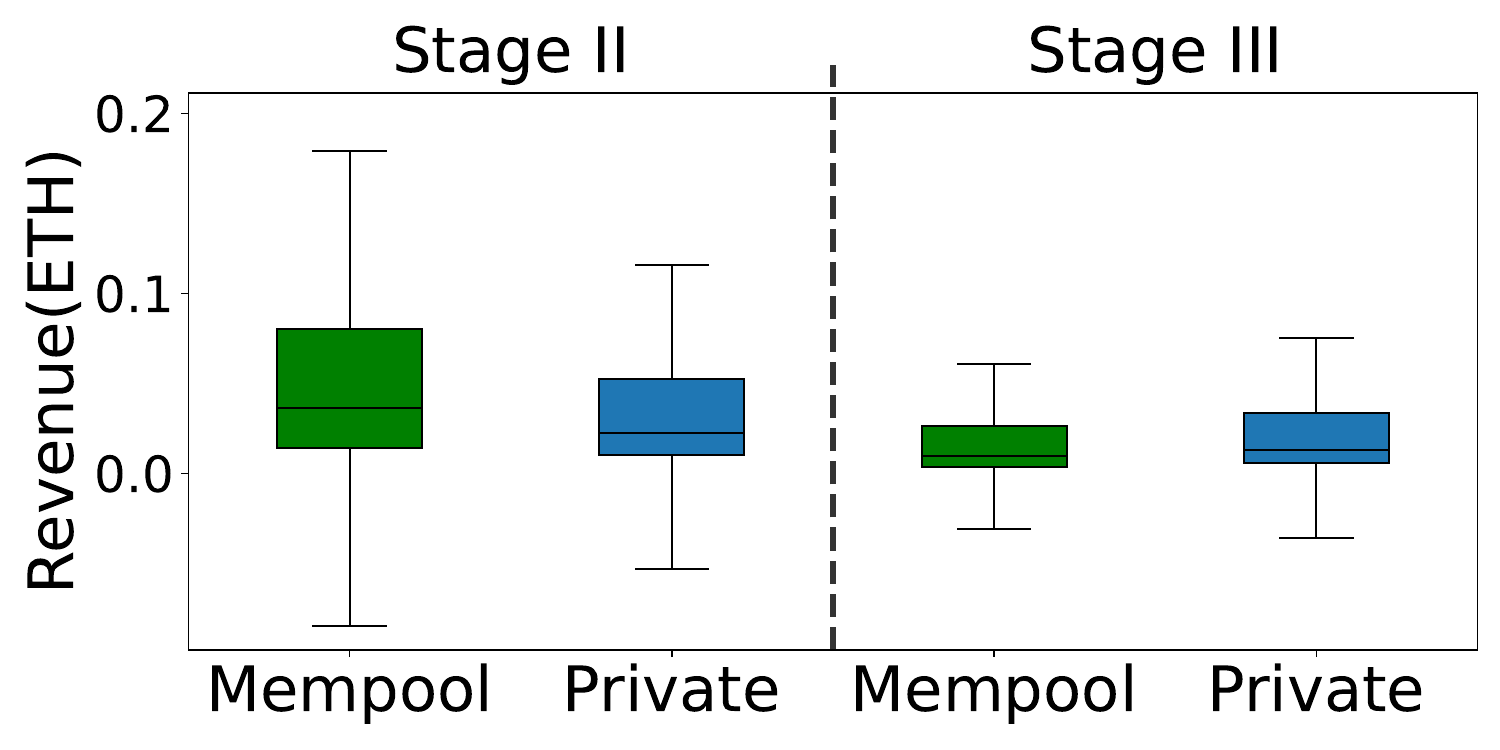}
        \vspace{-0.2in}
        \caption*{(a) Revenue}
    \end{minipage}\hfill
    \begin{minipage}{0.32\textwidth}
        \centering
        \includegraphics[width=\textwidth]{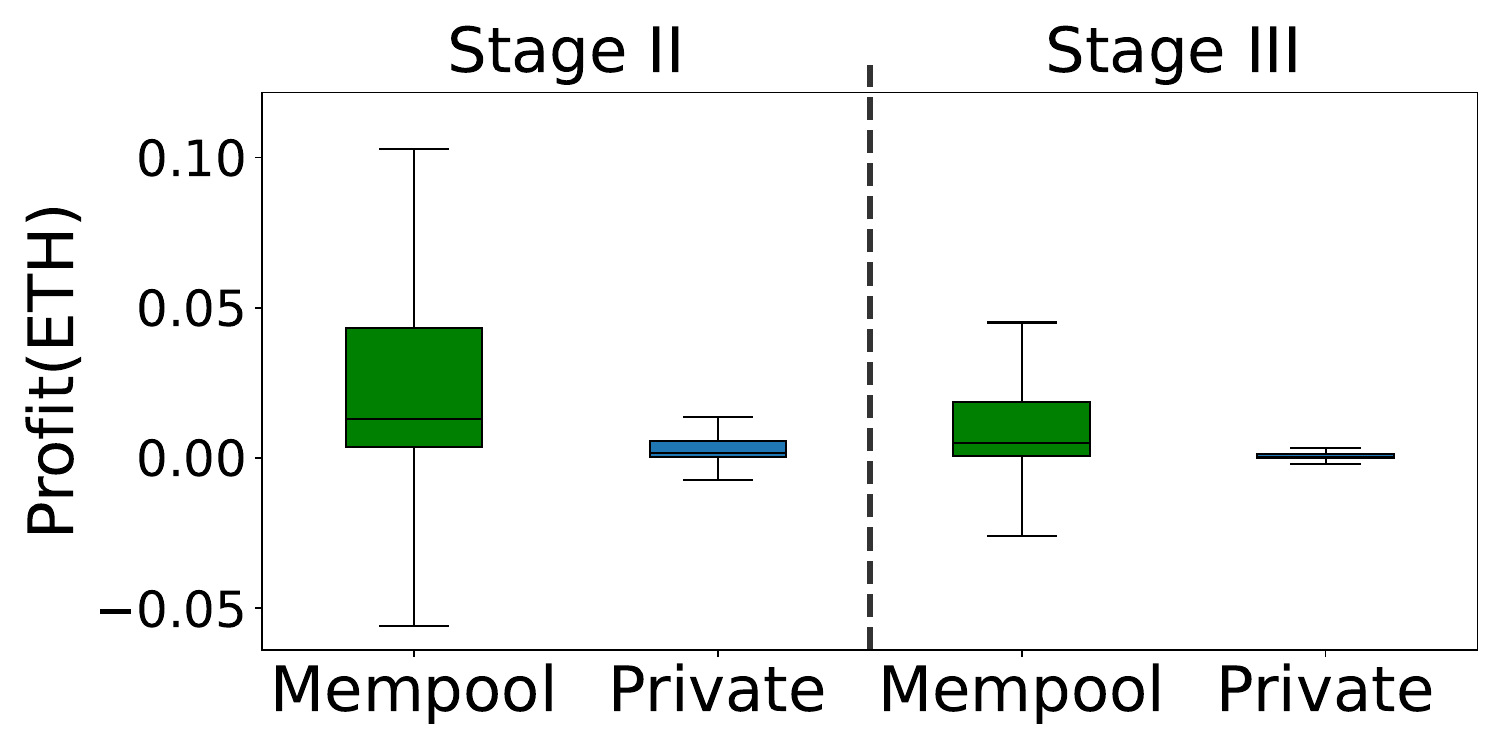}
        \vspace{-0.2in}
        \caption*{(b) Profit}
    \end{minipage}\hfill
    \begin{minipage}{0.32\textwidth}
        \centering
        \includegraphics[width=\textwidth]{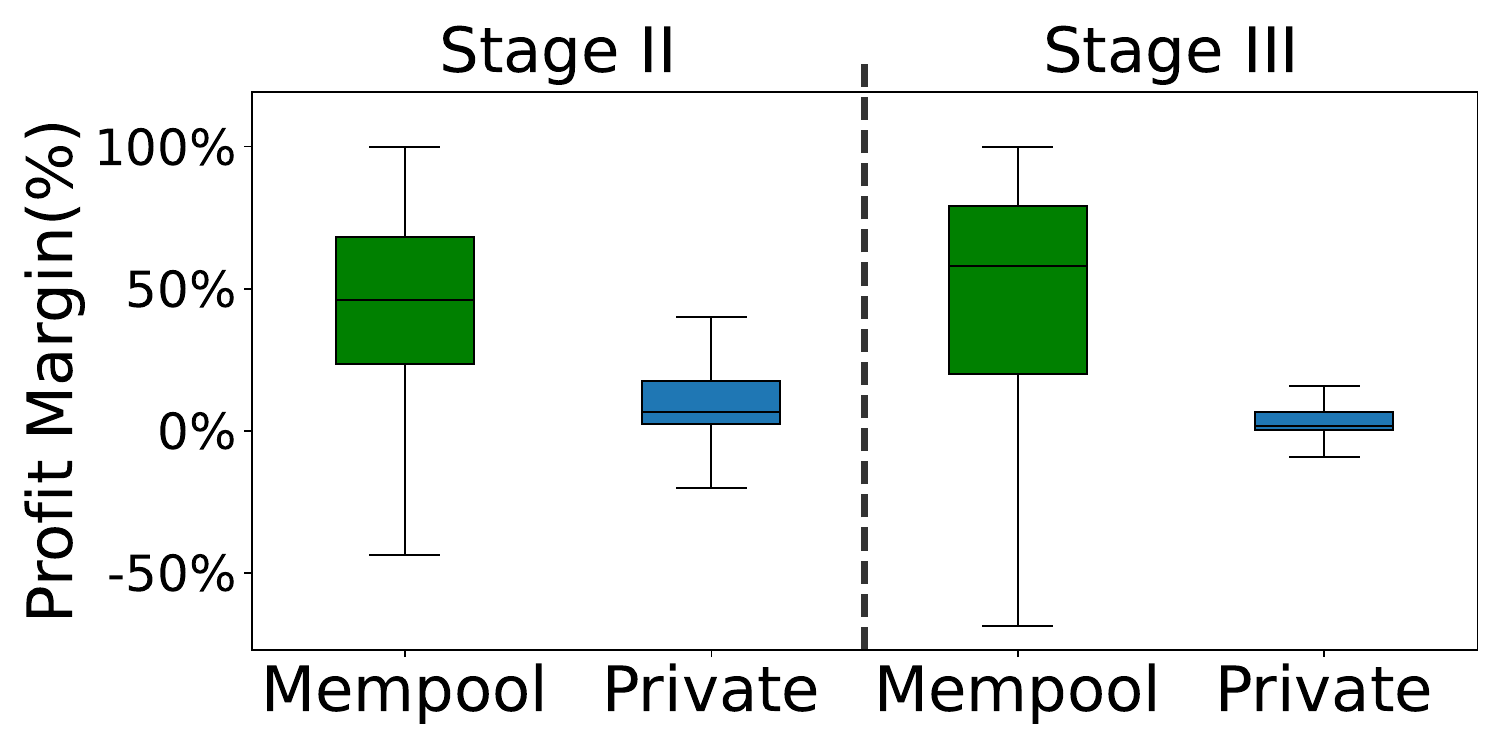}
        \vspace{-0.2in}
        \caption*{(c) Profit Margin}
    \end{minipage}
    \vspace{-0.1in}
    \caption{Revenue, profit, and profit margin of MEV activities in Stage II/III.}
    \vspace{-0.1in}
    \label{fig:Three_Boxplot}
\end{figure}

Fig.~\ref{fig:Three_Boxplot}(a) illustrates the revenue of these MEV activities.
Interestingly, we can observe that in Stage II, the median revenue of mempool MEV activities exceeded that of private MEV activities, while things started to go the other way in Stage III.
This suggests that in Stage III, MEV searchers prefer using private transactions to capture high-revenue opportunities rather than relying on mempool transactions.

Furthermore, Fig.~\ref{fig:Three_Boxplot}(b) illustrates the profit of MEV activities in these two phases. We can easily tell that in both stages, compared to private MEV activities, mempool MEV activities can get eight times higher profit.
Such a distinction exists even with the low proportion of mempool MEV activities (see Fig.~\ref{fig:Private_Pool}).
We speculate this is because transaction fees consume a significant portion of private transaction revenues, resulting in a decrease in profits.
To this end, despite MEV searchers being exposed to the risk of being front-ran, the allure of high profits continues to drive searchers to utilize mempool when extracting MEV.

Finally, we evaluate their profit margins, \textit{i.e.,} the ratio between profit and revenue, as shown in Fig.~\ref{fig:Three_Boxplot}(c).
As we can see, the profit margin of mempool MEV activities still exceeds that of private MEV activities, indicating that MEV searchers can obtain more profits within a certain amount of revenue.
In Stage II, the median profit margins for mempool and private MEV activities were 47.1\% and 6.7\%, respectively, while these two ratios changed to 66.6\% and 1.9\% in Stage III.
The gap between profit margins even widens, which we speculate the reasons are twofold: (1) private transaction architectures implement a bidding mechanism where MEV searchers compete by offering high transaction fees, thereby reducing their profit margins; and (2) for mempool MEV activities, since some block proposers might not be connected to private transaction architectures, or their block-building strategies might not prioritize the highest fee MEV activities, there is still a chance that MEV activities with lower fees can be proposed.
Moreover, this widening gap indicates that MEV profits are further skewed towards block proposers in Stage III, and the private transaction architecture did not democratize the distribution of MEV gains.

\textit{\textbf{Findings:}} Though the introduction of PBS incentivizes the adoption of private transaction architecture for MEV activities, the competition on transaction fee significantly lowers the profit, which exacerbates the centralization of profit gaining for block proposers.

\subsection{Adopted MEV Strategies}
\label{sec:Success Rate}
In Stage II and Stage III, MEV searchers begin to weigh the options between utilizing the mempool and private transaction architectures.
To this end, by discussing Stage II and Stage III together, we evaluate the shift of strategies based on the success rate and expected profit for MEV activities.

\noindent
\textbf{Success Rate.}
As Flashbots Relay in Stage II and Builders in Stage III verify the status of private transactions in MEV activities, reverted ones will not be included in blocks, allowing MEV searchers to avoid financial losses from transaction fees~\cite{failed_transaction}. Therefore, we assume the success rate of private MEV transactions is 100\%. However, for ordinary mempool transactions, failed transactions still require fees as they are executed regardless of success.
We focus on front-running arbitrages to evaluate the success rate of mempool MEV activities, whose reasons are twofold: (1) arbitrage represents the most prevalent form of MEV activities, and (2) failed sandwich attacks cannot be reliably detected as multiple transactions are required in a single attack. 
We use the method mentioned in \S\ref{sec:MEV_Identification} to detect successful front-running arbitrages. For failed front-running arbitrage, if a non-MEV transaction, after replayed at the top of the block, becomes an arbitrage, then it is classified as a failed front-running arbitrage.

\begin{figure}[t]
    \centering
    \begin{minipage}{0.49\textwidth}
        \centering
        \includegraphics[width=\textwidth]{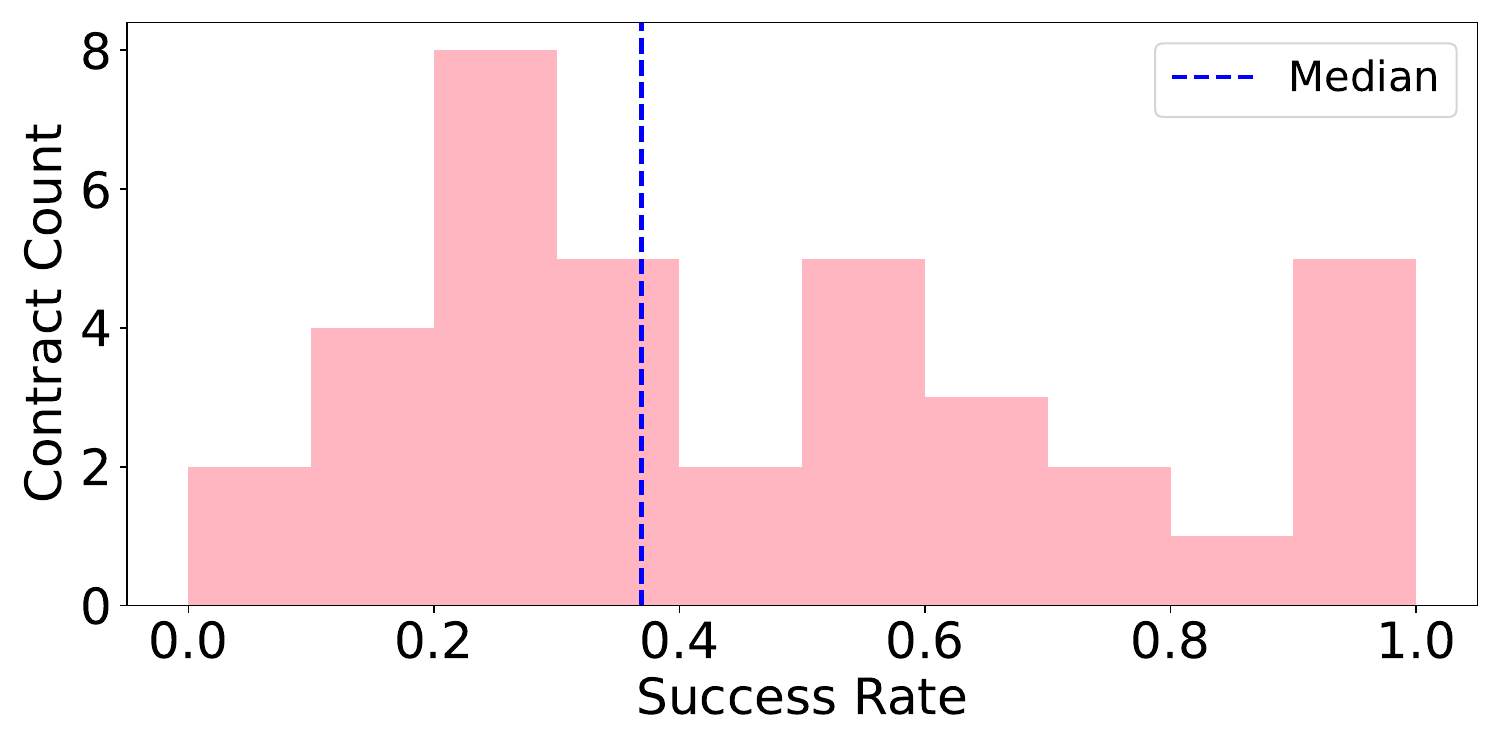}
        \caption*{(a) Success Rates (in mempool)}
    \end{minipage}\hfill
    \begin{minipage}{0.49\textwidth}
        \centering
        \includegraphics[width=\textwidth]{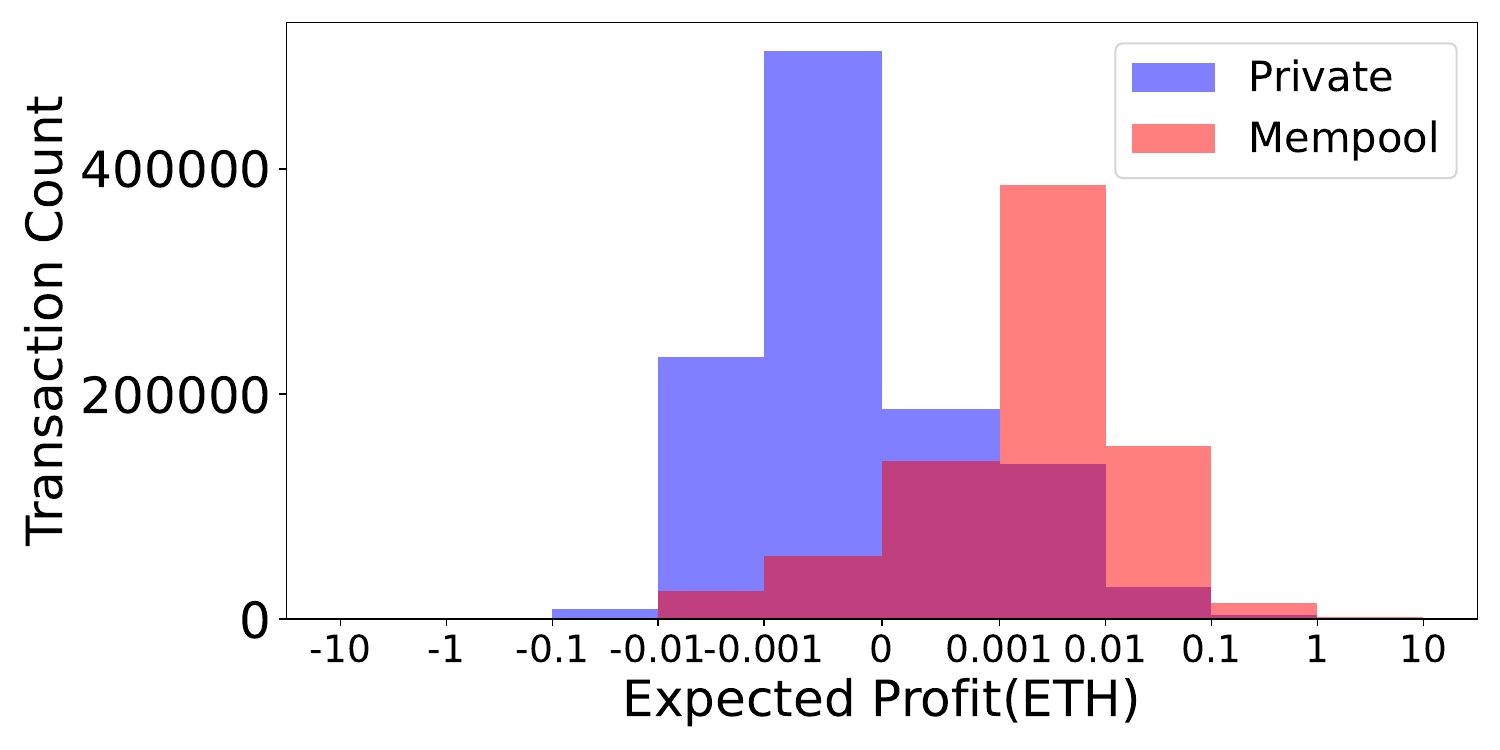}
        \caption*{(b) Expected Profit}
    \end{minipage}
    \caption{Distribution of success rates and expected profit for MEV searchers.}
    
    \label{fig:success_rate_expected_profit}
\end{figure}

We have replayed 20,496,511 transactions from 107 contracts that conduct more than 10,000 arbitrages. Among them, there are 779,055 successful mempool front-running arbitrages, 1,104,891 successful private front-running arbitrages and 675,065 failed mempool front-running arbitrages.
Fig.~\ref{fig:success_rate_expected_profit}(a) illustrates the distribution of contracts with different success rates. We can observe that the median success rate of top MEV searchers with mempool is below 40\%.
This suggests a high failure rate for mempool MEV activities, prompting many MEV searchers to switch to private transaction architecture. This shift explains the decline in mempool MEV activities in Stage II/III, as discussed in \S\ref{sec:rq2:tx-volume}.

\noindent
\textbf{Expected Profit.}
We introduce Expected Profit (\textit{EP}) to simulate the profit of MEV activities if they are broadcasted in mempool. $EP$ is calculated as:
$$EP = profit \times sr - gp \times fg \times (1 - sr)$$
, where $profit$, $sr$, $gp$, and $fg$ refer to the profit, the success rate, the gas price, and the consumed gas, respectively.
For 47 contracts that have initiated failed mempool arbitrages, $sr$ is calculated on contract basis and $fg$ is the average gas consumption of failed mempool arbitrages. 
For the remaining 60 contracts, $fg$ and $sr$ are calculated as the average of the ones of the 47 contracts.
In other words, if the expected profit is negative, the MEV searchers tend not to initiate it in mempool.
As shown in Fig.~\ref{fig:success_rate_expected_profit}(b), we observe that for 1,104,891 successful private front-running arbitrages, 747,798 (67.7\%) have a negative \textit{EP}. In contrast, out of 779,055 successful mempool arbitrages, only 82,077 (10.5\%) have a negative \textit{EP}, underlining the necessity of the private transaction architecture. Because around 70\% private front-running arbitrages will lead to financial losses to MEV searchers if they are broadcasted in the mempool.
That is, without private transaction architecture, smaller-yield MEV opportunities may not be extracted by MEV searchers.

\textbf{\textit{Findings:}} MEV activities through mempool suffer from a relatively low success rate (<40\%), thus the ones with little profits have to be extracted through the private transaction architecture, whose necessity is underlined.

\begin{tcolorbox}[colback=white, colframe=black, sharp corners, boxrule=0.2mm, width=\linewidth]
\textbf{Answer to RQ2: } 
After introducing private transaction architecture, MEV searchers gradually abandoned the mempool when conducting MEV activities. 
Moreover, the use of private transaction architecture avoids the financial losses due to the charge of transaction fees for failed mempool transactions, allowing MEV searchers to capture lower-profit MEV opportunities in today's highly competitive MEV environment.
\end{tcolorbox}

\section{RQ3: Demystifying Back-running in MEV}
\label{sec:RQ3}
According to McLaughlin~\cite{mclaughlin2023large}, back-running has gradually become the primary mechanism for arbitrages due to intensive competition among MEV searchers.
In this section, we delve deeper into the back-running to understand the characteristics and application of back-running arbitrages.

\subsection{Overall Results}
\label{sec:backrunning:overal}

\begin{figure}[t]
    \centering
    \begin{minipage}{0.49\textwidth}
        \centering
        \includegraphics[width=\textwidth]{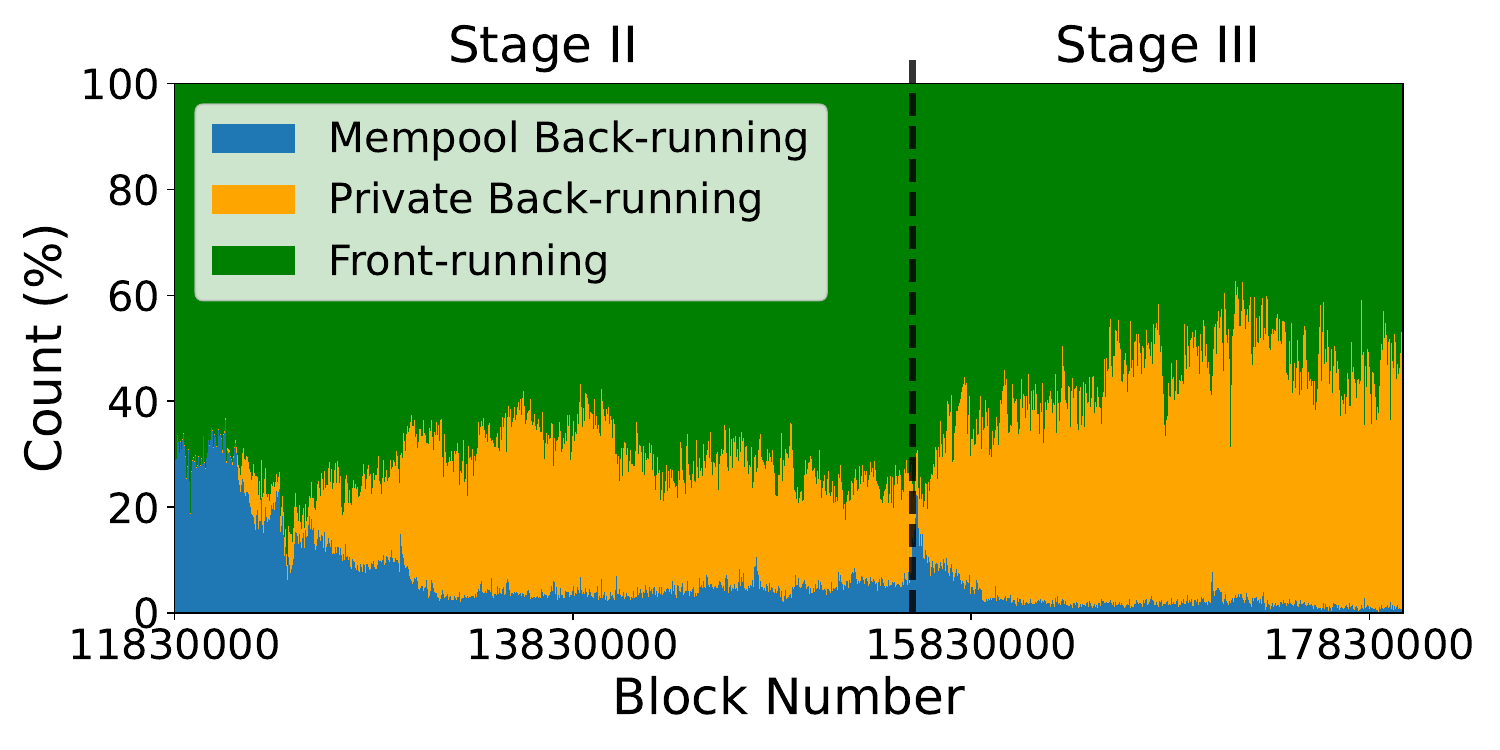}
        \vspace{-0.3in}
        \caption*{(a) Count}
    \end{minipage}\hfill
    \begin{minipage}{0.49\textwidth}
        \centering
        \includegraphics[width=\textwidth]{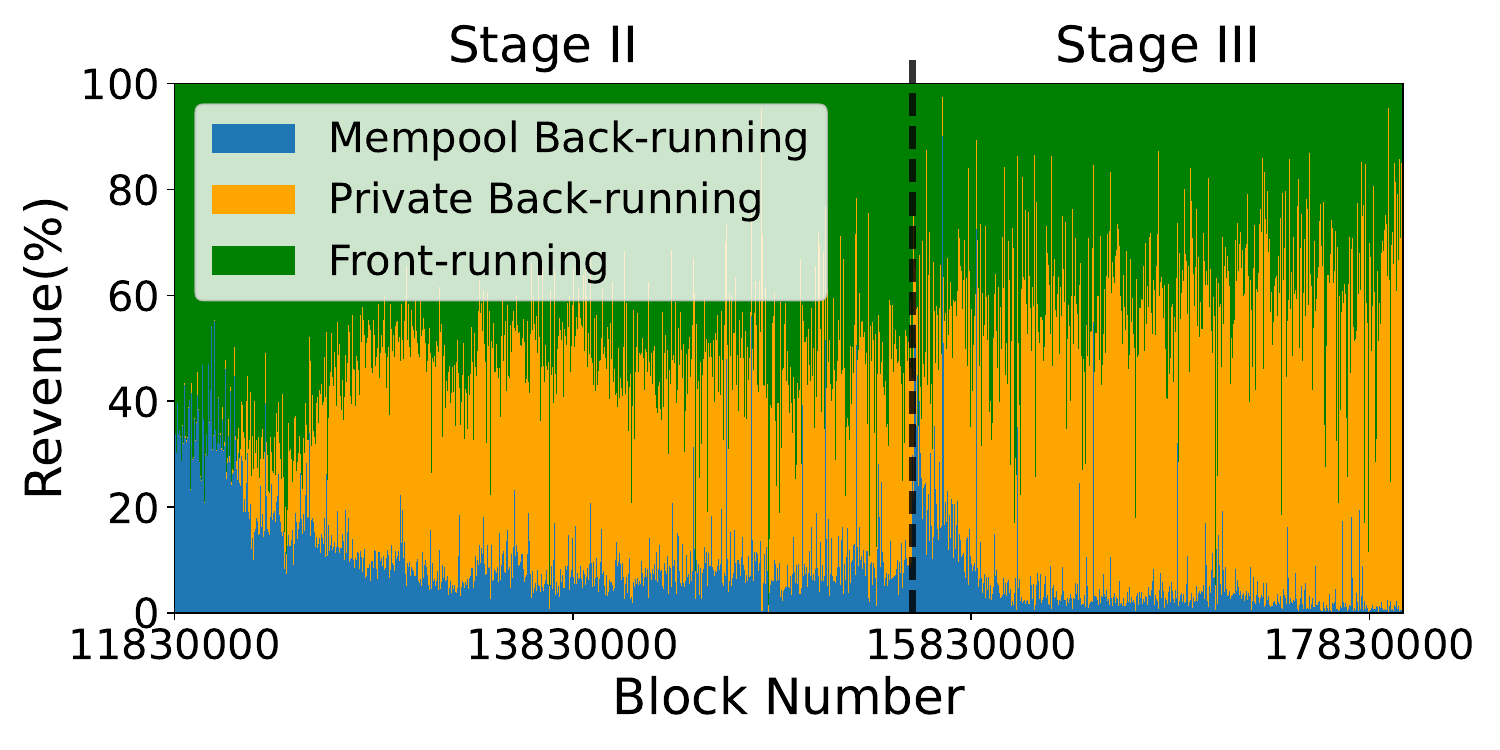}
        \vspace{-0.3in}
        \caption*{(b) Revenue}
    \end{minipage}
    \vspace{-0.1in}
    \caption{The distribution of different types of arbitrages in Stage II and III.}
    \label{fig:back-running-count-revenue}
\end{figure}

We utilize the method in \S\ref{sec:MEV_Identification} to identify back-running arbitrages.
Consequently, we identify 1,212,672 back-running arbitrages (879,282 private ones and 334,797 mempool ones) and 2,712,357 front-running arbitrages in Stage II and III.
Fig.~\ref{fig:back-running-count-revenue} illustrates the ratio among these three types in terms of number and revenue.
As we can see, back-running arbitrage is gradually becoming the main form of arbitrage. Since Stage III, it even experienced another great boom. There are two possible reasons.
First, as private transaction architecture is becoming prevalent, especially in Stage III, which provides transaction ordering services, \textit{e.g.,} Flashbots bundles, lowering the technical barrier to implement back-running arbitrages.
Second, the increasing number of MEV searchers leads to intensive competition. The MEV opportunities with high revenue are captured immediately in the same block, \textit{i.e.,} in the form of back-running arbitrage, rather than being left for the next block, which will be captured by front-running arbitrages.

\begin{table}[t]
\centering
\caption{Labels of target transaction contracts.}
\label{tab:backrunning_contract}
\resizebox{0.8\textwidth}{!}{
\begin{tabular}{@{}ccccccc@{}}
\toprule
\textbf{} & \textbf{Uniswap} & \textbf{MEV-Bot} & \textbf{SushiSwap} & \textbf{1inch} & \textbf{0x Protocol} & \textbf{Other} \\ 
\midrule
\textbf{\# Transaction} & 506,713  & 164,580 & 88,738 & 81,622 & 34,654 & 336,365 \\ 
\textbf{Proportion (\%)} & 41.8 & 13.6 & 7.3 & 6.7 & 2.9 & 27.7 \\
\bottomrule
\end{tabular}
}

\end{table}

As for the target transactions involved in these back-running arbitrages, we employ the identification method outlined in \S\ref{sec:eval:setting}.
We found that for 79.7\% cases, no other intermediate transactions are located between the target transaction and the back-running arbitrage one.
Table~\ref{tab:backrunning_contract} further shows the label of the receiver of all these target transactions with the help of Etherscan.
As we can see, 41.8\% of transactions are related to Uniswap. Out of them, 58.9\% are Uniswap V2 Router~\cite{uniswapv2router2021}. This suggests that public trading contracts generate a large number of back-running opportunities, which are targeted by MEV searchers.

\textbf{\textit{Findings:}} As the competition among MEV searchers intensifies, back-running arbitrage is playing an increasingly important role in both terms of number and revenue. Famous DEXes are typically considered targets for such MEV activities.

\subsection{Back-running Application\#1: MEV-Share \& MEV-Blocker} 
\label{sec:MEV-Share}

\begin{figure}[t]
\begin{minipage}{0.48\textwidth}
    \centering
    \includegraphics[width=\linewidth]{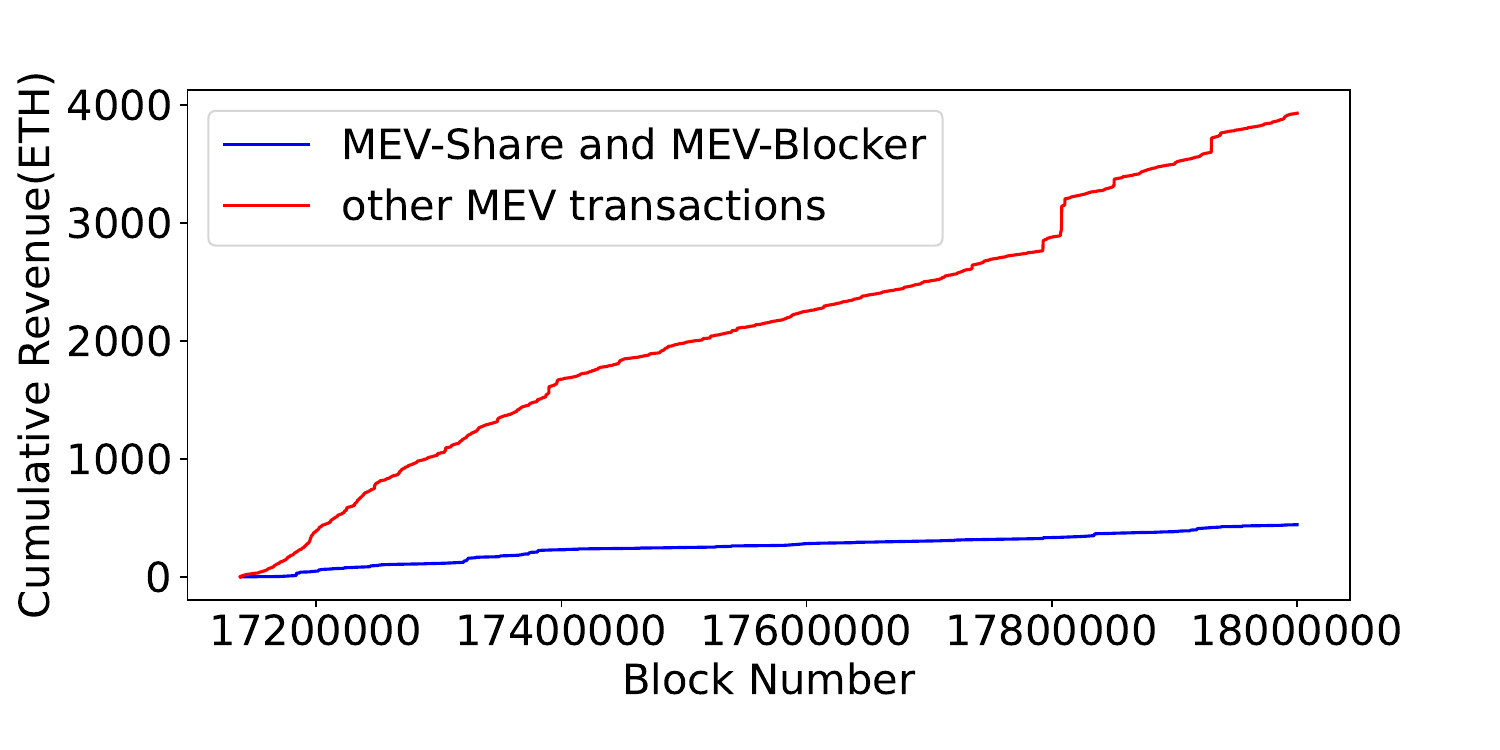}  
    \vspace{-0.3in}
    \caption{MEV searchers obtained revenue through back-running arbitrage.}
    \label{fig:mev_share_back_running_revenue}
\end{minipage}
\begin{minipage}{0.48\textwidth}
    \centering
    \captionof{table}{Median revenue distribution of MEV Share and MEV Blocker.}
    \resizebox{\textwidth}{!}{
    \begin{tabular}{@{}cccc@{}}
        \toprule
        \textbf{Service} & \textbf{Users} & \textbf{Searchers} & \textbf{Builders/Validators} \\
        \midrule
        MEV Share & 23.9\% & 44.2\% & 30.6\% \\
        MEV Blocker & 37.1\% & 19.8\% & 42.5\% \\
        \bottomrule
    \end{tabular}
    \label{tab:mev_share_port}
}
\end{minipage}
\end{figure}

We try to depict the characteristics of MEV-Share and MEV-Blocker, both of which are MEV-related services that emerged in Stage III.
Specifically, MEV-Share is a transaction service provided by Relays, which utilizes back-running techniques to refund a portion of profits back to the users whose transaction exposes MEV opportunities. This service not only increases Builders' profits but also returns profits to users.
Cow Protocol~\cite{cowprotocol} also provides a similar service, called MEV-Blocker~\cite{mevblocker}.
Transactions involved in both services are provided by the corresponding service providers~\cite{MEV-Sharedata,MEV-Blockerdata}.

We first analyze the revenue obtained by MEV-Share and MEV-Blocker, where we collected 589 and 8,422 transactions, respectively. Then, we parse all contracts in the \texttt{to} field of these targeted and back-run transactions and filter out all back-running arbitrages against them. In total, we identified 98,117 instances of back-running arbitrages, while only 6,109 of them are related to these two services.
Fig.~\ref{fig:mev_share_back_running_revenue} illustrates their revenues. As we can see, there are approximately 4,122.24 ETH extracted by MEV searchers in total, while only 441.76 ETH are captured by MEV searchers of MEV-Share and MEV-Blocker, accounting for a mere 10.7\%.
We conclude that these two services only protect a limited number of users, highlighting the growth potential of both services.

We then analyze the revenue distribution of back-running arbitrages conducted by these two services. As shown in Table~\ref{tab:mev_share_port}, MEV searchers, \textit{i.e.,} the ones who find the MEV opportunities, receive the largest proportion of revenue in MEV-Share, while they are builders and validators in MEV-Blocker.
Notably, users receive the smallest share in MEV-Share. In contrast, in MEV-Blocker, the users' revenue share is comparable to that of builders and validators.
This shows that, compared to MEV-Share, MEV-Blocker distributes more benefits to users, builders, and validators, encouraging everyone to use its services.

\textbf{\textit{Findings:}} Only 10.7\% of revenue obtained by MEV is protected by conducting back-running arbitrages from MEV-Share and MEV-Blocker, reflecting their significant growth potential. Moreover, MEV-Blocker would return more profits, which incentivizes them to adopt its service.

\subsection{Back-running Application\#2: Builder MEV Extraction} 
\label{sec:Builder MEV Extraction}

\begin{table*}[t]
    \centering
    \caption{Participation of builders in back-running arbitrages in Stage III.}
    \label{tab:MEV_Builders_back-running}
    \resizebox{\linewidth}{!}{
    \begin{tabular}{cccc}
        \toprule
        \textbf{Builder Address} & \textbf{MEV Searcher Address} & \textbf{\#Transactions} & \textbf{Revenue (ETH)} \\
        \midrule
        0x3bee5122e2a2fbe11287aafb0cb918e22abb5436 & 0x9dd864d39fbfdf7648402746263e451cd4f36af0 & 590 & 81.0\\
        0x3b64216ad1a58f61538b4fa1b27327675ab7ed67 & 0xb0bababe78a9be0810fadf99dd2ed31ed12568be & 381 & 40.5\\
        0x229b8325bb9ac04602898b7e8989998710235d5f & 0x6c6b87d44d239b3750bf9badce26a9a0a3d2364e & 141 & 85.0\\
        \bottomrule
    \end{tabular}
    }
\end{table*}

In Stage III, Builders can provide private transaction services through their owned APIs. As Builders can arbitrarily reorder received transactions, it allows them to perform MEV~\cite{eigenphi2023}.
Therefore, Builders can capture a significant portion of the MEV revenue without facing competition.
This deviates from the PBS's vision~\cite{pbs}: \textit{ Builders might have done sophisticated MEV extraction, but the reward for it goes to the Validators.}
In this section, we try to filter out if this type of situation happens regularly.

We filter out all back-running arbitrages in private transaction pools between Sep. 2022 to Aug. 2023. A total of 54,273 such transactions are identified.
Heuristically, if more than 50\% of the MEV activities in blocks constructed by a builder originate from an identical MEV searcher, and more than 50\% of the MEV searcher's activities are packed by the builder, we assume they are collusive.
As shown in Table~\ref{tab:MEV_Builders_back-running}, we have filtered out three such groups. We calculate the revenue of their MEV activities and find that the median profit margins for them are 80.6\%, 46.1\%, and 44.7\%, respectively. Compared to the results mentioned in \S\ref{sec:MEV profit}, we can conclude that these MEV searchers do not undergo severe competition, which is most likely due to the collusion.
Interestingly, the second group, its builder is named Boba Builder, and the MEV searcher contract is also named Boba. According to its website~\cite{bobabuilder}, we have not seen any statement indicating that it engages in back-running arbitrages for private transactions. 

\textbf{\textit{Findings:}} In Stage III, PBS enables Builders to conduct back-running arbitrages quietly by reordering transactions. Heuristically, we have observed that nearly 620K USD are involved in such hidden activities, indicating a need for better regulatory protocols within PBS.

\begin{tcolorbox}[colback=white, colframe=black, sharp corners, boxrule=0.2mm, width=\linewidth]
\textbf{Answer to RQ3:}
Due to increasingly fierce competition among MEV searchers, back-running began to become the mainstream MEV strategy.
Additionally, though MEV-Share and MEV-Blocker can return part of users' profits through back-running, they are not widely adopted. We also observe that Builders are sneakily using back-running to conduct MEV from private transactions, yielding significant profits, which contradicts the vision of PBS, urging strong regulatory measures for PBS.
\end{tcolorbox}

\section{Limitations}
We acknowledge certain limitations of our method.
First, in terms of data collection, although we have compiled the most extensive dataset of MEV transactions for arbitrage and sandwich attacks, we still rely on previous methods (mentioned in \S\ref{sec:data_overview}) to detect private transactions. This reliance may lead to false positives.
Second, regarding evaluating our algorithms for identifying arbitrages and sandwich attacks, our benchmark for manually judging whether a transaction is one of these types is as follows. For MEV activities, they must include addresses that show a profitable balance change in tokens, and the addresses should not be irrelevant ones (such as the Uniswap V2 Router ~\cite{uniswapv2router2021}, which is not controlled by any traders). The transactions must also exhibit the characteristics of each type of MEV: sandwich attacks, where the attacking transactions are positioned before and after the victim's transaction, and arbitrage, which involves a cyclic swap pattern. Although this manual inspection process is applicable to most MEV transactions of these two types, we cannot perform a precise evaluation of our method due to the lack of access to the ground truth.

\section{Related Work}
Our work is related to the following three categories.  

\noindent
\textbf{MEV identification.}
Most research~\cite{qin2022quantifying,wang2022cyclic,weintraub2022flash,piet2022extracting,torres2021frontrunner} are based on heuristic methods and are only identified for a few popular Dexes. The advantage of them is their simplicity, but the drawback is the limitation on specific types of transactions.
Our work improves upon this heuristic method by expanding the range of Dexes searched and refining the rules, enabling the identification of more subtypes of transactions.
Mclaughlin et al.~\cite{mclaughlin2023large} utilizes the heuristic method proposed by Wu et al.~\cite{wu2021defiranger}, relying solely on ERC-20 transfer events to identify DEX addresses and detect arbitrage transactions, thereby expanding the scope of Dexes searched.
Recent work by Li et al.~\cite{li2023demystifying} utilizes transaction aggregation, detecting MEV transactions on a Flashbot bundle basis, which can uncover more types of MEV transactions. 

\noindent
\textbf{Private transaction architecture.}
Before the PBS update, several studies~\cite{weintraub2022flash,capponi2022evolution,piet2022extracting,lyu2022empirical} focus on the relationship between private transaction architecture and MEV. After the PBS update, other studies~\cite{wahrstatter2023time,heimbach2023ethereum} delve into the characteristics of the PBS update and its impact on MEV. Wahrstätter et al.~\cite{wahrstatter2023time} investigated the impact of the time taken by builders to submit blocks on their block construction profits, analyzing whether the builder strategy involves waiting to accept more MEV transactions or submitting blocks in advance. Our research focuses on the impact of private transaction architectures on the profits and strategies of MEV searchers.

\noindent
\textbf{MEV applications.}
Due to the potential losses incurred by traders due to MEV, some applications have been designed to mitigate MEV. 
For example, some solutions aim to prevent losses caused by front-running through the design of DEX~\cite{zhou2021a2mm,ciampi2022fairmm,mcmenamin2022fairtradex,baum2021p2dex,cowprotocol}.  
Other solutions involve redesigning the order of transactions~\cite{khalil2019tex,bentov2019tesseract,malkhi2022maximal}. 
In our study, although back-running transactions do not directly impact traders' profits, the majority of traders do not profit from the MEV generated by them. Despite MEV-Share and MEV-Blocker being an effective solution, they have not been widely adopted, indicating that the MEV problem introduced by Ethereum has not been adequately addressed and remains a challenge in terms of fair distribution of profits.

\section{Conclusions}
In this work, we innovatively proposed a profitability identification algorithm, based on which, we further designed two new algorithms to identify arbitrages and sandwich attacks in MEV ecosystem.
On our collected largest ever dataset, compared to state-of-the-art methods, our algorithms can identify MEV activities with lower false positive/negative rate, at most 2.4\%.
Our experimental results uncover some interesting findings: 1) Except for well-known DeFi projects, emerging meme tokens are also widely taken as targets for MEV activities; 2) The emergence of private transaction architecture is necessary. Though it reduces the obtained profits for MEV searchers, it allows them to capture lower-profit MEV opportunities; and 3) Back-running began to replace front-running in arbitrages. However, we observe that some applications of back-running arbitrages are not fully adopted, and Builders may maliciously take advantage of back-running arbitrages.
We believe our findings can shed the light on the future direction.

\bibliographystyle{splncs04}
\bibliography{cite}

\end{document}